# Методы панорамной визуализации и цифрового анализа теплофизических полей. Обзор.

Автор: И.А. Знаменская

МГУ им. М.В. Ломоносова

ORCID: 0000-0001-6362-9496   znamen@phys.msu.ru

**Аннотация**

Работа посвящена обзору современных достижений в области регистрации, обработки, анализа динамических процессов в жидкостях, газах, плазме, многофазных средах, реализующихся в природе и технике. Автор обзора рассматривает как физические основы визуализации потоков, так и основы современных технологий цифровой обработки зарегистрированных изображений. Краткий анализ истории развития методов панорамной визуализации и теплофизических полей перекрывает период полутора столетий. В работах последнего десятилетия в фокусе - методы компьютерной обработки, средства, технологии анализа и распознавания изображений панорамных теплофизических полей, позволяющие получать количественную информацию о потоках. Обзор содержит анализ публикаций, описывающих основные современные методы визуализации потоков: методы основанные на явлении рефракции; использовании электролюминисценции; цифровое трассирование (particle image velocimetry), визуализации поверхностных течений (бароиндикаторные и термоиндикаторные покрытия, жидкие кристаллы; масляные покрытия); Особое внимание уделено методам, использующим кросс-корреляционные алгоритмы обработки изображений. Рассматриваются методы, где этот алгоритм является основным: цифровое трассирование (PIV), теневой фоновый метод (ТФМ), беззасевные теневые методы, термографическое PIV, измерение скорости в вязких покрытиях, микро, томографические модификации PIV и др. Затрагивается актуальная проблема анализа цифровых данных панорамного эксперимента - проблема больших данных. Приводятся примеры использования машинного обучения при анализе больших массивов данных теневой съемки. Рассмотрены некоторые примеры визуализации данных численного моделирования, имитирующих эксперимент.

**Ключевые слова**: панорамная визуализация, цифровой анализ изображений, оптические методы, поле скорости, градиент плотности, распознавание разрывов, теневой фоновый метод, компьютерная обработка.

# 1. ВВЕДЕНИЕ.

Визуализация движущихся сред в природе и в технических устройствах как способ фиксации и передачи информации о различных динамических процессах существует несколько столетий. Понимание сложных нестационарных процессов, происходящих в потоке газа, жидкости, плазмы возможно при наблюдении всей картины течения среды в целом. В связи с этим, панорамная визуализация поля течения среды является важнейшим способом получения информации о таких потоках при экспериментальных исследованиях в теплофизике, механике газа, жидкости, плазмы.

Принято считать, что в эпоху Возрождения появилась научная визуализация потоков – динамические процессы и явления теплофизики, механики жидкости, газа фиксировались в графических изображениях. Исследования потоков газа и жидкости помогало гению итальянского Возрождения Леонардо да Винчи продумывать проекты создания средств передвижения по воздуху, воде, суше. С эпохи Возрождения визуализация развивалась преимущественно в двух видах - как искусство и как наука. Вновь открытые физические принципы регистрации изображений использовалось одновременно и на новом витке научной визуализации, и в искусстве. Например, изобретение фотографии первоначально использовалось в основном в быту; лишь в конце XIX века появилась научная фотография. В 1839 году в Париже на собрании с участием членов Академии наук было объявлено, что Дагер открыл способ проявлять и закреплять фотографические изображения. После этого события началось развитие научной визуализации.

Панорамная визуализация теплофизических полей включает двумерное, трехмерное, четырехмерное представление результатов измерений и цифрового анализа исследуемых полей газодинамических, теплофизических параметров, а также и результатов численных расчетов динамических процессов. Панорамная визуализация потоков востребована во многих естественных науках, - теплофизике, геофизике, медицине, физике плазмы, биологии, психологии и др.

А.Теплер впервые предложил так называемый метод свилей (разновидность теневого метода) для изучения оптических неоднородностей среды и наблюдал ударную волну от электрической искры (1867 г.). Э.Мах предложил несколько диагностических методик. Он получил в 80-е годы 19 века первые теневые фотографии головной ударной волны перед пулей, летящей со сверхзвуковой скоростью, позже получил интерферограммы потока. [1]. Французский естествоиспытатель Ж.Э. Марей впервые применил многокадровую фотографию для регистрации движений и, построив аэродинамическую трубу, визуализировал системой дымов обтекание препятствий и переход к турбулентности. [2].

В течение последних двух веков развитие панорамной визуализации теплофизических, гидродинамических полей определялось преимущественно следующими пятью факторами:

Прогрессом в создании регистрирующих материалов;

Прогрессом в создании регистрирующей изображения аппаратуры;

3. Созданием новых источников генерации (и передачи) зондирующего электромагнитного излучения;

4. Развитием носителей визуальной информации (способов хранения).

5. Развитием цифровых методов анализа и обработки результатов визуализации (начиная с 80-годов 20 века).

Регистрирующие материалы эволюционировали от твердых носителей (бумага, фотопластинки, фотопленка) до цифровых матриц. Регистрирующая аппаратура соответственно - от прямого наблюдения глазами, далее наблюдения с использованием камер-обскур, оптики. С середины XIX века для научной визуализации стала использоваться фотоаппаратура, затем кинокамеры, видеокамеры, барабанные высокоскоростные камеры. В середине XX века появились электронно-оптические приборы. Источники света – также постоянно эволюционируют: использовалось – солнечное излучение, свечи, лампочки, электрические разряды, лазеры, кумулятивные, светодиодные источники и др. Современные источники излучают в широком диапазоне длин волн и длительности импульса.

В наступившую цифровую эпоху идет постоянное обновление технологий изготовления специализированных камер для регистрации процессов в потоках; идет конкурентная борьба фирм-изготовителей за увеличение пространственно-временных, спектральных характеристик камер.

Методы визуализации структуры и параметров потоков основаны на физических свойствах электромагнитного излучения и его взаимодействий со средами, включая различные физические процессы:

- Рассеяние;
- Рефракция;
- Поглощение
- Отражение
- Интерференция
- Дисперсия
- Люминесценция
- Эффекты нелинейной оптики

Перечисленные процессы включают проявления свойств света, которые могут быть проинтерпретированы в рамках геометрической, волновой, квантовой моделей. Сегодня для грамотного использования экспериментальных методов визуализации в теплофизике, механике и анализа результатов панорамных экспериментов необходимо быть специалистом по физической оптике; теплофизике и механике, а также по цифровым методам обработки и анализа научных изображений. В последние годы в связи с непрерывно увеличивающимся объемом визуальных данных панорамного эксперимента начался процесс подключения машинного обучения и использования искусственного интеллекта для их анализа.

Изображения полей течений в теплофизике - основной источник информации о распределении параметров среды, конфигурациях и динамике структур в потоке - вихрей, слабых возмущений, сильных разрывов - ударных волн, контактных разрывов, о линиях тока, турбулентных структурах и др. Цель визуализации теплофизических потоков - качественное и количественное исследование, описание новых явлений и закономерностей в потоках в различных средах, а также в их демонстрации. Второй важной целью накопления данных визуализации динамических теплофизических полей является обеспечение бенчмарков для тестирования программ и алгоритмов численного моделирования в теплофизике.

За последние 10-20 лет произошел стремительный переход на цифровые технологии регистрации, обработки, анализа динамических процессов в жидкостях, газах, плазме, многофазных средах, реализующихся в природе и технике. Публикуются (а в последние 5 лет – в основном размещаются в интернете) в большом количестве соответствующие цифровые издания – журналы, сборники, монографии. Классические монографии и альбомы визуализации течений 80-х - 2000х годов [3-8] сосредоточены на описании оптических схем, регистрирующей аппаратуры, анализе изображений тестовых потоков. Получаемая в 20 веке при визуализации информация (в доцифровую эпоху) носила преимущественно качественный характер. В последние десятилетия число зарубежных обзорных публикаций по теме многократно увеличилось – например - [9-11] и др. Среди монографий и обзоров последних 20 лет по панорамным методам на русском языке следует отметить [12-18].

В работах последнего десятилетия в фокусе - методы компьютерной обработки, средства, технологии анализа и распознавания изображений теплофизических полей, позволяющие получать количественную информацию о потоках. Одной из целей получения визуальной информации о динамических процессах в потоках жидкости, газа, плазмы – создание баз экспериментальных данных для верификации программ трехмерных расчетов [19] Актуальные задачи развития экспериментальной базы для верификации CFD кодов при использовании в атомной энергетике. М. А. Большухин и др. Труды НГТУ им. Р.Е. Алексеева 2013 Т 2(99) стр. 117-125 ]

С появлением мощных компьютеров возникла возможность проводить вычислительные эксперименты, основанные на численном решении уравнений, используемых в математической модели исследуемого физического явления или процесса. Системы научной визуализации позволяют представить результаты вычислений и сравнивать их с данными эксперимента [20-21].

Актуальная проблема анализа данных панорамного теплофизического эксперимента - проблема больших данных. В процессе экспериментов сегодня накапливается огромное количество цифровой информации, полученной при видеосъемке на цифровые камеры, тепловизоры, и т.д. При работе с большими данными подключается проблема машинного обучения при анализе больших массивов данных (в данном случае – изображений исследуемых потоков). Пока весьма немного работ посвящены данной проблеме, но их число растет быстро.

Панорамная визуализация теплофизических полей включает ряд этапов: 1.Визуализация потока. 2. Регистрация изображения изменяющейся области (поля) потока. 3.Цифровая обработка изображения (фильма) 4.Получение количественной информации. 5.Получение полей физического параметра. 6.Анализ и интерпретация, физическая модель.

В настоящее время развитие методов панорамной визуализации и цифрового анализа теплофизических полей определяется во многом внедрением в экспериментальную практику современных программных и электронных средств для ввода в ЭВМ изображений, полученных при визуализации, и их цифровой обработки. С помощью специализированного программного обеспечения обрабатываются изображения как в теплофизике, так и механике сплошных сред, медицине, геофизике, биологии. При этом ставятся сходные базовые задачи: уменьшение зашумленности исходного изображения, выделение элементов структуры исследуемых объектов, сохранение полученных результатов в удобном для дальнейшей работы и представления виде.

Особенно много количественной информации о параметрах теплофизических полей получено за последние года благодаря использованию кросс-корреляционных алгоритмов обработки изображений. Этот алгоритм является основным в модификациях теневого фонового метода (ТФМ), анемометрии по изображениям частиц (PIV – Particle Image Velocimetry), включая микро, стерео, томографические модификациях, беззасевную анемометрию, термографическом PIV, при измерениях скорости в вязких покрытиях, и др.

## 2.МЕТОДЫ, ОСНОВАННЫЕ НА РЕФРАКЦИИ СВЕТА

Для визуализации многих видов оптически прозрачных потоков в теплофизике и динамике сплошных сред используются методы визуализации, основанные на явлении отклонения света при его прохождении через неоднородности плотности прозрачной среды: теневой метод, шлирен – метод, интерферометрию, а также их модификации. Оптический показатель преломления среды n равен отношению скорости света в среде к скорости света в вакууме и связан с локальной плотностью среды формулой Лоренц-Лорентца, которая для газов имеет вид:

$$\frac{n-1}{\rho} = k,$$

где k-постоянная величина, (для воздуха равная 0,22635 см$^3$/г).

Теневой метод обнаружения неоднородностей плотности в газе был предложен в 1867 году немецким ученым Теплером. Шлирен - метод визуализации иногда так и называется – метод Теплера [22]. Если поток газа неоднороден, то оптический показатель преломления среды в исследуемой области потока зависит от координат (x, y, z). При просвечивании области течения с переменной плотностью луч, распространяющийся параллельно оси z и проходящий через неоднородность, отклоняется от первоначального направления распространения на угол α

$$\alpha \cong \int_0^L \frac{\partial}{\partial x} \ln n(x, y, z) dz \ldots$$

Основным недостатком теневых методов является то, что все изменения плотности суммируются вдоль направления распространения луча зондирующего излучения и таким образом регистрируется интегральное значение изменения плотности. Поэтому теневые и интерференционные методы с успехом применяются для визуализации двумерных, а также некоторых осесимметричных газодинамических течений. При визуализации поля течения газа теневым методом изменение освещенности пропорционально степени изменения градиента плотности газа. При наличии в потоке сильных градиентов плотности – (в частности - поверхностей разрыва) происходят дополнительные отклонения луча на поверхности разрыва. Теневое изображение ударной волны представляет собой темную полосу со стороны набегающего потока, сменяющуюся яркой светлой полосой, интенсивность которой постепенно уменьшается (Рис. 1). При лазерном зондировании на теневых изображениях возможно появление дифракционных полос в области разрывов.

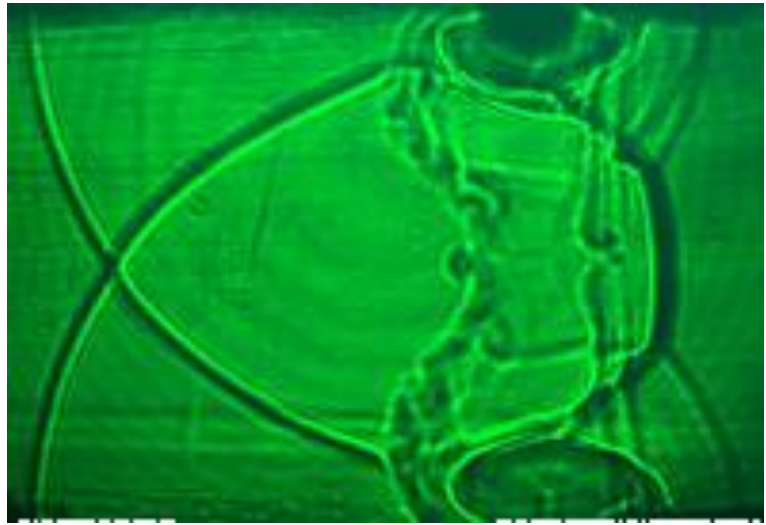

Рис.1. Теневое изображение квазидвумерного процесса взаимодействия ударной волны с областью импульсного поверхностного разряда с лазерной подсветкой.

В работе [23] теневым методом исследованы четыре сценария диффузионного горения круглой микроструи водорода в зависимости от скорости ее истечения. В [24] теневым и шлирен методами исследована струя газа, истекающая из сопла.

Шлирен-метод визуализации (или метод Теплера) – это усложненный теневой метод. Основной принцип действия шлирен-системы состоит в том, что часть света, отклоненного при прохождении через неоднородность плотности газа, задерживается кромкой ножа, установленного в фокальной плоскости пучка, прошедшего через исследуемую область. На экране, вследствие этого, освещенность соответствующих частей изображения уменьшится или увеличится в зависимости от того, куда направлено отклонение. Изменение освещенности в точке, сопряженной с неоднородностью, определяется величиной угла отклонения луча, фокусным расстоянием второго объектива и размером источника света. В варианте шлирен-метода, называемом методом нити, перекрываются, наоборот, неотклоненные лучи. При визуализации поля течения газа шлирен-методом изменение освещенности пропорционально градиенту плотности газа в исследуемой области в направлении, перпендикулярном кромке ножа, а не степени изменения градиента плотности, как в теневом методе. Шлирен - методом лучше визуализируются вихри, волны разрежения; теневым методом более точно регистрируется положение разрывов.

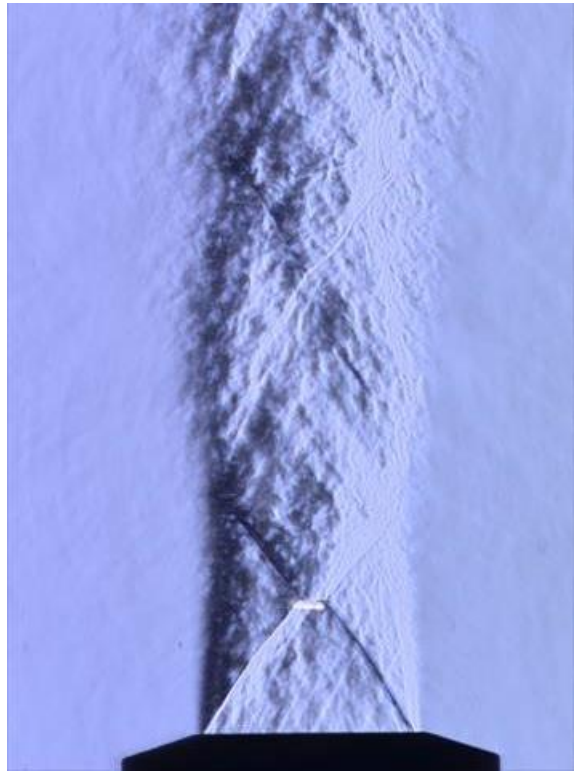

Рис.2. Шлирен регистрация течения в сверхзвуковой струе; оптический разряд в сверхзвуковом потоке.

Визуализацию шлирен-методом течения, близкого к осесимметричному, можно рассмотреть на примере изображения струи газа (рис.2), истекающей из сопла [24].

Усложненная модификация шлирен - метода – цветной шлирен-метод [25, 26]. Существует несколько способов получения цветной теневой картины. Чаще всего в приемной части оптической системы или в фокальной плоскости коллиматора устанавливается многоцветная диафрагма. Цветные шлирен-изображения потока дают больше информации о поле течения, чем черно-белые, поскольку на них вместо одной плотности почернения изменяются сразу три компоненты, а именно цвет, насыщенность цвета и яркость изображения. Глаз способен различать больше цветных оттенков, чем оттенков серого, поэтому шлирен - системы с цветным изображением дают, как правило, больше качественной информации. К преимуществам цветного теневого метода относится то, что детали спектра обтекания на цветной картине легко отличить от границ моделей и других объектов, затеняющих картину течения.

Оптическая интерферометрия используется в основном для количественных исследований плотности двумерных прозрачных потоков. При интерферометрической регистрации поля течения картина распределения полос интенсивности света отражает распределение показателя преломления среды. Интерференционная картина представляет собой систему полос, расстояние между максимумами которых при заданной длине волны $\lambda$ определяется углом схождения интерферирующих волн. Введение оптической неоднородности в одно из плеч интерферометра изменяет оптическую длину пути соответствующего пучка по сравнению с невозмущенным и вызывает смещение интерференционных полос. Для измерений плотности в потоках жидкости и газа используют интерферометры Майкельсона, Жамена, Маха — Цендера, Фабри-Перо. При исследовании двумерных областей течения (в которых параметры газа по направлению прямолинейного распространения луча не меняются) интерферометр Фабри-Перо обладает повышенной чувствительностью. При сдвиге интерференционных полос на величину δ (в долях расстояний между полосами) значение плотности меняется на величину $\Delta\rho/\rho=\delta\lambda/kL$, где $L$ — геометрическая длина пути луча в неоднородности k-постоянная Гладстона-Дэйла. Интерферометр Маха — Цендера, обладает некоторыми преимуществами при исследованиях в газовой динамике. В этом интерферометре с помощью системы полупрозрачных пластин и зеркал осуществляется амплитудное деление световой волны на две и последующее наложение волн, прошедших различными оптическими путями. Для осесимметричных неоднородностей смещение интерференционных полос δ (x) в сечении, перпендикулярном оси симметрии течения, связано с изменением плотности решением интегрального уравнения Абеля.

$$\delta(r,x) = \frac{2k}{\lambda} \int_{r}^{R(x)} \frac{\rho(y,x) - \rho_0}{\sqrt{y^2 - r^2}} y dy$$

где $R$ — граница неоднородности в сечении; r,x — цилиндрические координаты, $\rho_0$ — плотность невозмущенной среды.

Большую сложность при анализе интерферограмм представляет определение изменения плотности при переходе через поверхность разрыва, особенно для плоских течений. Трудно либо невозможно установить соответствие полос на интерферограммах при переходе через ударную волну при съемке с монохроматическим источником света. Для решения этой проблемы существуют специальные методы интерферометрических измерений.

Теневая визуализация взаимодействий ударных волн с препятствиями позволила установить закономерности перестройки структуры течений с разрывами. В частности, при отражении ударной волны от твердой поверхности было обнаружено существование нескольких типов конфигураций, реализующихся при различных условиях отражения. Впервые это явление было изучено Э. Махом, исследования ведутся и сейчас, поскольку вопрос о критических углах перехода от одного типа отражения к другому не решен однозначно. Важным является анализ поля плотности в окрестности точки пересечения ударной волны с поверхностью с точки зрения установления значения критического угла $\alpha_{кр}$. Поэтому исследования поля течения потока при отражении ударной волны ведутся как теневыми, так и интерференционными методами.

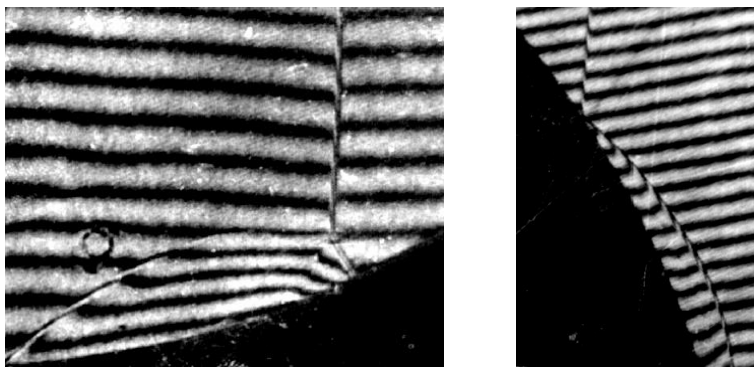

Рис.3. Отражение ударной волны от вогнутой и выпуклой цилиндрической поверхности.

На рис.3 приведены две интерферограммы (Гвоздева Л.Г., и Сысоев Н.Н., 1970-е гг.) двумерного нестационарного потока - отражения ударной волны от вогнутой выпуклой цилиндрической поверхности. В работе [27] описаны результаты интерферометрии неоднородности газодинамических окон лазера со сверхзвуковым потоком активной среды с одновременной регистрацией двух интерферограмм с взаимноортогональным направлением сдвига. Апробировано программное обеспечение обработки данных.

С помощью методики зондирующей микроинтерферометрии были получены экспериментальные зависимости радиального распределения электронной плотности от времени для фемтосекундной лазерной микроплазмы оптического пробоя газов (воздух, азот, аргон и гелий) при различных давлениях (от 1 до 10 атм) [28].

На рис.4 приведена интерферограмма, визуализирующая конвективные структуры и фазовый переход в вертикальном слое пресной воды [29].

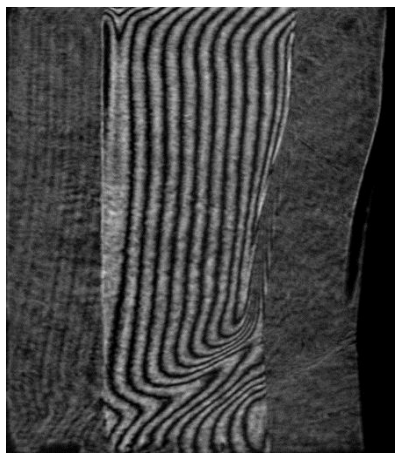

Рис.4. Сдвиговая интерферограмма.

В работе [30] с использованием интерферометра Маха-Цендера фазовые сдвиги получены из интерферограмм за время 1000 нс методом быстрого преобразования Фурье, а затем значения показателя преломления выведены с помощью инверсии Абеля.

Однако, в целом количество работ по интерферометрии теплофизических полей заметно уменьшилось за последние десятилетия. В какой-то мере на замену интерферометрии пришел теневой фоновый метод.

Теневой и шлирен методы существуют с 19 века, однако за последние десятилетия произошло несколько важных усовершенствований [31]. Теневая съемка- по-прежнему основной метод панорамной визуализации для течений с разрывами и турбулентностью, в том числе – течений жидкости. На рис.5 приведены теневые фотографии типичных событий на поверхности воды, приводящих к генерации капель [32].

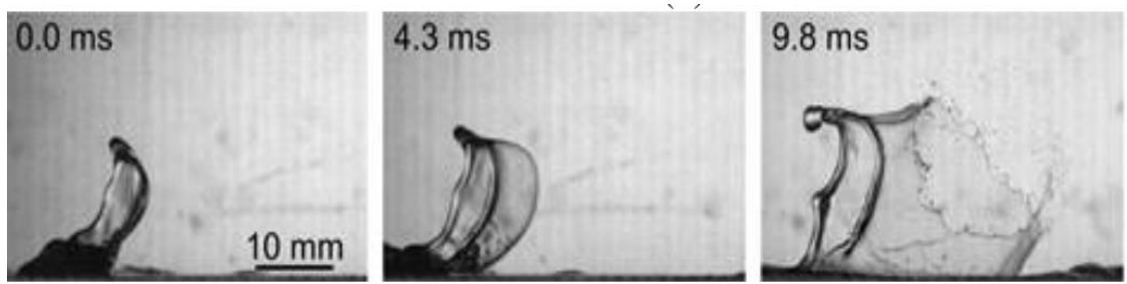

Рис.5 Формирование и разрыв структур при образованиикапель.

Применение высокоскоростных цифровых камер позволило изучать длительные высокоскоростные нестационарные процессы с частотой съемки более миллиона кадров в секунду. На рисунке 6 приведены 4 последовательных теневых кадра встречного движения ударных волн, инициированных поверхностным разрядом наносекундной длительности при съемке высокоскоростной цифровой камерой (124 тыс. кадров/с).

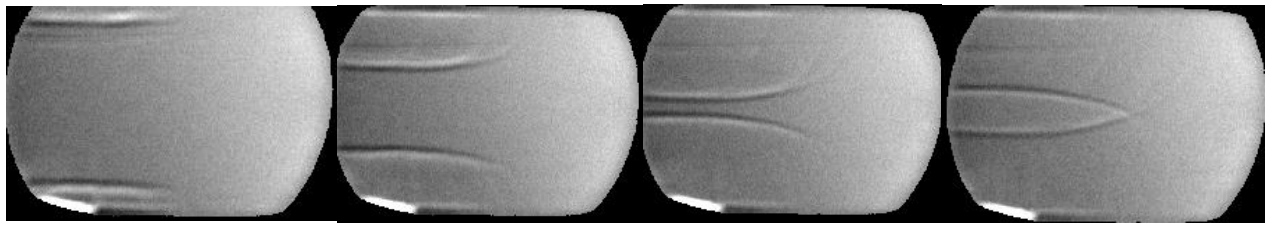

Рис. 6. Теневые кадры высокоскоростной съемки ударных волн от поверхностного разряда.

# 3.ВИЗУАЛИЗАЦИЯ ПОТОКОВ ГАЗОВЫМ РАЗРЯДОМ.

Визуализация газодинамических потоков электрическим разрядом основана на том, что газовая среда является при ее ионизации источником излучения (явление электролюминесценции), интенсивность которого связана с локальными газодинамическими параметрами. При горении стационарного объемного разряда в потоке газа пространственные неоднородности плотности среды ведут к перераспределению интенсивности излучения плазмы разряда. Этот эффект достаточно часто используется для визуализации сверхзвуковых течений в аэродинамических трубах при низких давлениях [33]. Достоинство этого метода - простота аппаратурного оформления и возможность наблюдения динамических изменений картины течения в ходе эксперимента. Неоднородность плотности приводит к перераспределению тока за счет сильной зависимости концентрации электронов и проводимости от величины коэффициента ионизации. Коэффициент ионизации является нелинейной функцией отношения напряженности поля к концентрации нейтральных частиц E/N. В зависимости от отношения величины линейного масштаба, характеризующего градиент концентрации частиц к различным характерным физическим масштабам (дебаевскому радиусу, длине свободного пробега электронов и т. д.) возможны различные физические эффекты, связанные с градиентами плотности газа. Локальная интенсивность излучения плазмы разряда I в некоторый момент времени:

*I(x,y,z,)*= f (W, U, i, N, P, T, d, Se)

Где W- электрическая энергия, подводимая к разрядной области, U - напряжение между электродами, i - ток разряда, N (или ρ)- плотность частиц, P - давление , T - температура, d- расстояние между электродами в мм, Se – параметр, характеризующий геометрию электродов. Начиная с 50-годов в аэродинамических трубах ЦАГИ при помощи стационарного высоковольтного разряда проводились исследования потоков около различных моделей [34-37]. Были визуализированы, в частности, стационарные ударные волны в течении около модели конуса на цилиндрической державке при разных углах атаки. Визуализируемые элементы потока – вихревые жгуты срыва потока вдоль поверхности конуса. Проводились исследования вихревых течений около других изолированных тел вращения, тел вращения с крылом; визуализирована головная ударная волна около цилиндра с полусферической головной частью.

В работах японских исследователей визуализация сверхзвуковых течений с 90-х годов производится также методом стационарного электрического разряда. Визуализировались ударноволновые конфигурации около различных моделей при числе Маха до 10. [38-39]. Стационарный разряд реализовывался между линейным и точечным электродами. При прохождении разряда по участку, где есть скачок, ударная волна визуализируется как темная область. Были получены изображения конфигурации ударных волн в поперечном сечении относительно модели. Для увеличения участка потока, который возможно визуализировать, вводились дополнительные точечные электроды.

В Индии велись работы по визуализации сверхзвуковых течений около моделей: (затупленный конус, затупленный конус с шипом и диском. [40-41]). Использовался, так же, как в предыдущем случае, метод визуализации стационарным электрическим разрядом. Были визуализированы элементы потока около модели: головная ударная волна, точка присоединения потока, ножка Маха. Проводились численные расчеты для сравнения с полученными экспериментальными данными.

На данный момент не так много исследований проводится с использованием визуализации газодинамических течений методом электрического разряда. В основном визуализация ведется постоянным электрическим разрядом [42, 43]. На рис.7 приведено интегральное изображение сверхзвукового течения разреженного азота при подсветке поперечным тлеющим разрядом.

У фронта ударной волны при горении разряда в сверхзвуковом стационарном потоке формируются слои с пространственным зарядом, в которых изменяются напряженность электрического поля, концентрация заряженных частиц и проводимость газа, причем характер изменения величин существенно зависит от полярности поля.

Известно, что горение разряда в потоке может привести к нагреву газа, существенным изменениям структуры, формы разрывов, увеличению ударного слоя, усилению слабых возмущений и ряду других эффектов, в том числе и ведущих к качественным изменениям структуры течения. Таким образом, никак нельзя рассматривать визуализацию потока стационарным газовым разрядом как бесконтактный метод.

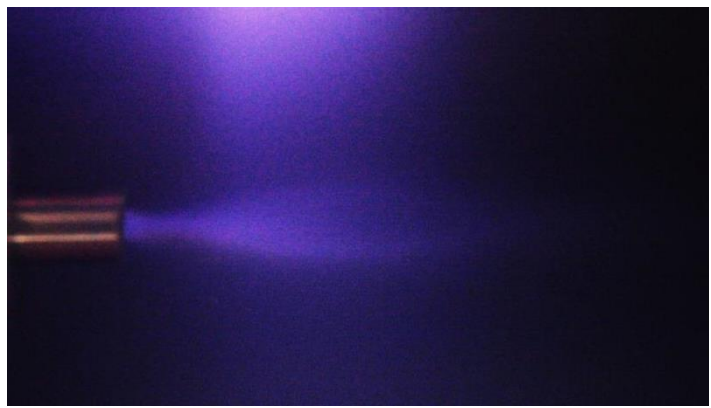

Рис.7. Визуализация тлеющим разрядом газовой струи из сверхзвукового сопла.

Некоторые недостатки визуализации газовым разрядом устраняются при использовании импульсного, исходно пространственно однородного разряда с временем горения, много меньшим, чем характерные времена газодинамических взаимодействий. При исследованиях на аэродинамических трубах эти времена – десятки и сотни микросекунд, при исследованиях быстропротекающих газодинамических процессов, включая нестационарные взаимодействия ударных волн – единицы микросекунд. При создании неравновесной пространственно однородной области потока эффективен импульсный объемный разряд с предыонизацией ультрафиолетовым излучением от плазменных листов [44-47]. Использование такого разряда обеспечивает минимальное время развития пробоя, диффузный характер свечения на начальных этапах развития разряда, исключают возможность возникновения в разрядном объеме неоднородностей, способных неконтролируемо повлиять на газодинамическое течение. Правильная прямоугольная форма разрядной области позволяет мгновенно ионизовать участок газодинамического потока, в частности, течения в ударных, аэродинамических трубах. На рисунках (7-9) представлены результаты визуализации нестационарных и стационарных газодинамических потоков с использованием такого разряда (МАИ, МГУ, 2000-2020гг). Визуализировались течения газа в ударной трубе 48х24 мм, имеющей рабочую камеру специальной конструкции того же сечения. Две стенки камеры – кварцевые окна; верхняя и нижняя – плоские плазменные электроды протяженностью 100 мм, поджигаемые в заданный момент процесса; одновременно поджигается объемный разряд. Время свечения ионизованного потока воздуха в рабочем диапазоне давлений составляет 150÷200 нс. При интегральной регистрации свечения поля течения "экспозиция" является мгновенной с точки зрения газодинамического временного масштаба. За время экспозиции - высвечивания элементов течения - при максимальных скоростях движения ударных волн 2000 м/с волна перемещается на 0,4 мм, при средних скоростях плоской ударной волны около 1000 м/с смещение составляет 0,2 мм. Возмущения, движущиеся с дозвуковой скоростью, "размазываются" за время экспозиции на сотые доли миллиметра. На рисунке 8а приведено изображение поля течения около затупленного цилиндра диаметром 9 мм при числе Маха сверхзвукового обтекания М=1,5 (искусственные цвета). Визуализированы головная ударная волна, висячие скачки, отраженный скачок. На фото .8б. поток за проходящей ударной волной (слева) формирует зону отрыва в донной части сферы.

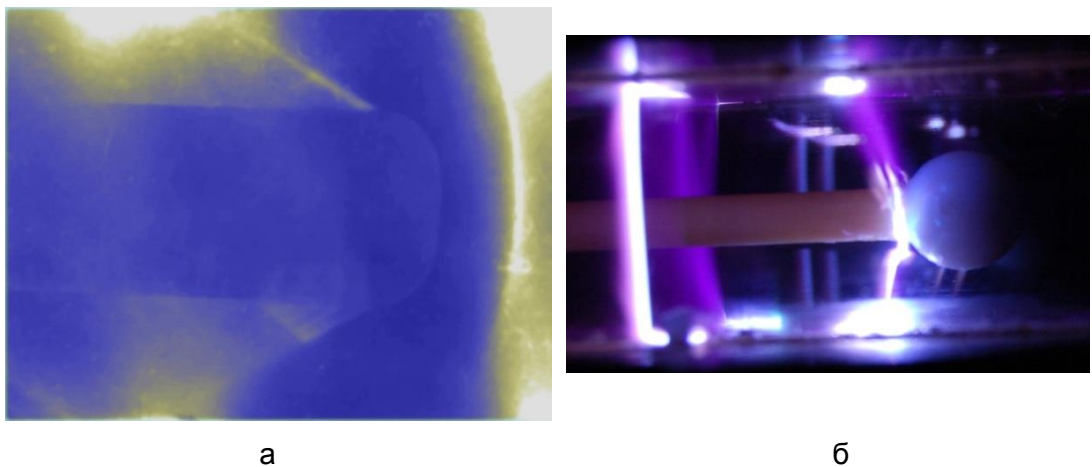

а                  б

Рис.8. Визуализация импульсным объемным разрядом стационарного течения около затупленного цилиндра и нестационарного обтекания шара.

В отличие от оптических методов, требующих зондирования объекта, данный метод позволяет визуализировать поток через одно окно рабочей камеры. С другой стороны, можно получать одновременно несколько изображений через оба окна рабочей камеры, при этом общий угол обзора может достигать 200° - 250° [44].

Визуализация неоднородностей с использованием электролюминесценции позволяет исследовать трехмерные структуры потока. Теневые методы могут дать изображение боковой проекции потока с регистрацией интегральных характеристик в направлении просвечивания, перекрывание моделью оптического пути исключает зондирование центральной (осевой) области. На рис. 9 приведены два изображения потока около осесимметричной модели, соединяющей в себе основные геометрические тела вращения, которая часто становится объектом исследования, являясь элементом многих конструкций в аэродинамике. Модель представляет собой конус с углом полураствора 10°, который крепится на цилиндре, затем диаметр цилиндра уступом увеличивается. На фото рисунка 9 приведено изображение сверхзвукового течения около модели, целиком помещенной в разрядный промежуток. Зарегистрирован момент формирования обтекания модели после прохода ударной волны с числом Маха М=2.8. Структура течения полностью визуализирована на двух изображениях, зарегистрированным через противоположные окна камеры.

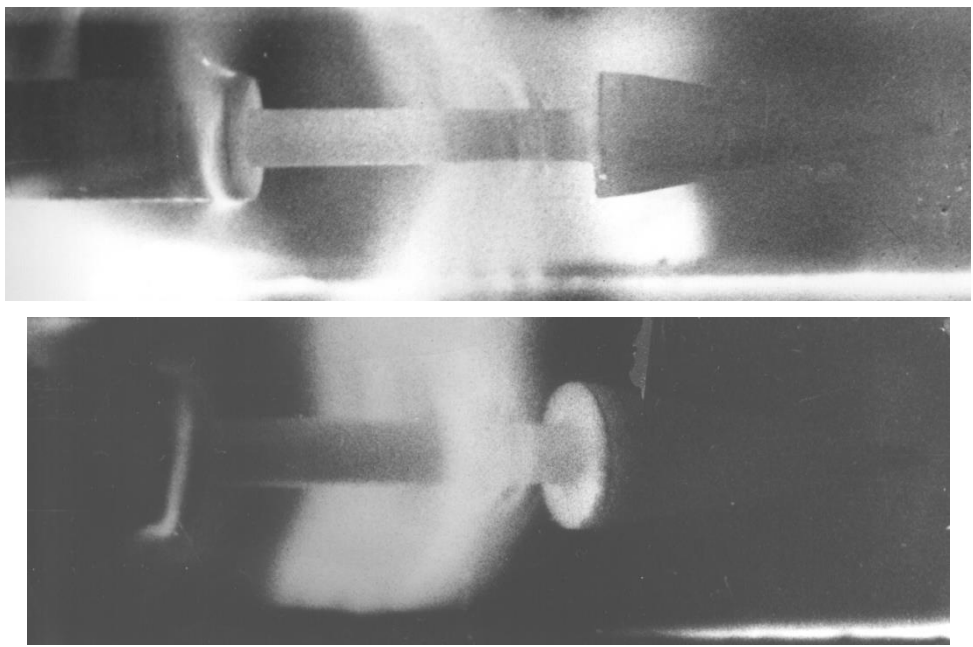

Рис. 9 Двухракурсная визуализация потока около осесимметричной модели.

Метод визуализации импульсным объемным разрядом позволил изучить динамику трехмерного вихревого кольца в донной области при дифракции ударной волны на конусе с уступом [45]. Также визуализировались нестационарные квазидвумерные течения в канале с уступами, возникающее в канале за падающей ударной волной. На верхней и нижней стенках канала от стекла до стекла располагались прямоугольные препятствия 2х6х48 мм [46,47]. На рисунке 10 импульсный объемный разряд визуализирует огибание плоской волной в канале препятствия на нижней стенке и цилиндрический вихрь в зоне отрыва. Объемный разряд перераспределяется в зону низкого давления перед ударной волной и в зону отрыва за уступом.

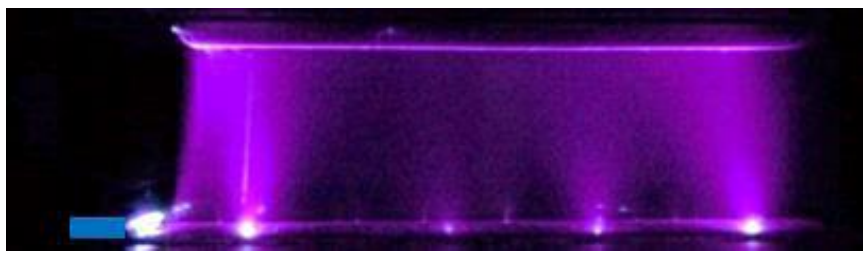

Рис.10. Свечение импульсного объемного разряда с предионизацией при дифракции ударной волны на препятствии на нижней стенке.

# 4. ПРОБЛЕМА БОЛЬШИХ ДАННЫХ В АНАЛИЗЕ ДАННЫХ ПАНОРАМНОЙ ВИЗУАЛИЗАЦИИ.

Актуальная проблема анализа данных панорамного эксперимента - проблема больших данных. В процессе экспериментов сегодня накапливается огромное количество цифровой информации, полученной при видеосъемке на цифровые камеры, тепловизоры, и т.д. Развитие цифровых технологий приводит к многократному увеличению объема визуальной информации о параметрах теплофизических полей, возникают большие массивы цифровых данных, которые не представляется возможным обработать вручную.

Так, современные видеофильмы, регистрирующие эволюцию турбулентных течений жидкости на основе теневых методов, трассирования, термографии, требуют обработки и квалифицированного анализа. Неизбежен переход на другой уровень анализа данных.

Возникает необходимость автоматизировать процесс обработки и анализа больших данных - экспериментальных изображений теплофизических полей, используя различные подходы, в том числе - методы машинного зрения и обучения, используя глубокое обучение, свёрточные нейронные сети (англ. convolutional neural network, CNN). Пока весьма немного работ посвящены данной проблеме, но их число растет быстро. Работа в данном направлении начинается в отдельных научных лабораториях. Создается программное обеспечение (ПО) для распознавания структурных элементов течений в газах, жидкостях и плазме. Для тестирования ПО и обучения нейронных сетей используются массивы изображений различных течений, зарегистрированные теневым, шлирен, PIV методами.

Сегодня наиболее перспективные подходы к решению этих проблем основаны на цифровой обработке изображений с различными алгоритмами обнаружения границ и распознавания объектов [48] для идентификации сложных структур потоков. Для обнаружения ударной волны подходят различные алгоритмы обнаружения краев изображения. Различными исследователями было показано, что алгоритм Canny [51] лучше всего подходит для обработки теневых и шлирен-изображений. Авторы [48] создали программное обеспечение для обнаружения и отслеживания ударных волн. Улучшенные с помощью выделения границ на шлирен-изображениях процесса смешивания в сопле были проанализированы в [52]. Обработка изображений выполнялась по алгоритму Canny. Проводятся работы по улучшению качества изображений с помощью удаления шумов, вычитания фоновых изображений различными методами. В [53] показано, что вычитание фонового изображения в частотном представлении (выполняемое с помощью быстрого преобразования Фурье), позволяет добиться наилучшего качества. Также перспективными для цифровой обработки экспериментальных изображений методами компьютерного зрения являются: методы сегментация изображений с использованием таких алгоритмов как k-средних, минимизации энергии и др.; методы выявления признаков (англ. feature detection) (SURF, LESS, HOG и др.).

В настоящее время активно развивается тема машинного обучения для гидродинамики. Достаточно подробный обзор исследований по этой теме приведен в ([54]. Также быстро развивается подход, основанный на машинном обучении, для идентификации структур потоков на шлирен-изображениях. В работе [55] предложена система классификации и обработки шлирен-изображений объектов в аэродинамической трубе. Система смогла извлечь из изображений три параметра: угол преломления головной ударной волны, разницу в интенсивности линии и среднюю ширину линии. Основываясь на этих данных, система вычисляет скорость потока вблизи модели. Нейронные сети могут извлекать не только ударные волны, но и любые другие элементы потока. Например, авторы [56] успешно применили нейронную сеть для классификации вихревых следов за профилем крыла. Нейронные сети начинают использоваться для предсказания [57] или реконструкции [58] динамики течений. Предлагаются новые, основанные на физических принципах, способы вычисления функции потерь (loss function) [58]. Глубокое машинное обучение позволяет моделировать турбулентность и другие газодинамические системы больших размерностей [59].

Для исследования эволюции газодинамического течения на протяжении 6-10 миллисекунд в работах [60-61] использовались результаты количественного анализа высокоскоростной теневой съемки течений в ударной трубе. При помощи машинного зрения и обучения [61] [62] было разработано три программы обработки теневых и шлирен-изображений. Первая работает на алгоритмах машинного зрения, вторая использует обученную нами свёрточную нейронную сеть для распознавания и автоматического отслеживания различных структур течений, третья программа использует метод кросс-корреляции для оценки скорости течений по смещению турбулентных структур и взвешенных в потоке частиц – по аналогии с методом PIV (англ. Particle Image Velocimetry). Для выделения границ использовался алгоритм Кэнни (англ. Canny edge detection), который оказался наиболее эффективным для анализированных теневых и шлирен-изображений. Алгоритм поиска угла косого скачка уплотнения включает в себя выделение границ, поиск уравнений прямых (скачков уплотнения) с помощью преобразования Хафа, фильтрация найденных прямых по длине, углу и положению. Определен момент перехода к дозвуковому режиму. Динамика плоских ударных волн автоматически измерялась с помощью сверточной нейронной сети. Скорость распознавания объектов на изображениях (ударных волн, частиц-трассеров в потоке) нейронной сетью составила 15 к / с. Были автоматически измерены и построены зависимости скорости падающей, отраженной УВ от времени. Исследована динамика цуга псевдоскачков (англ. shock train) в канале. Использование алгоритмов машинного зрения и обучения позволило ускорить обработку и анализ больших массивов экспериментальных цифровых изображений (Рис. 11) и полностью автоматизировать этот процесс. На ручную обработку одной съемки течения в ударной трубе (около 1000 кадров при скорости съемки 150 000 кадров / с) мог уйти целый рабочий день. Разработанное ПО решает эту задачу за одну-две минуты. Таким образом, было значительно ускорено получение новой физической информации.

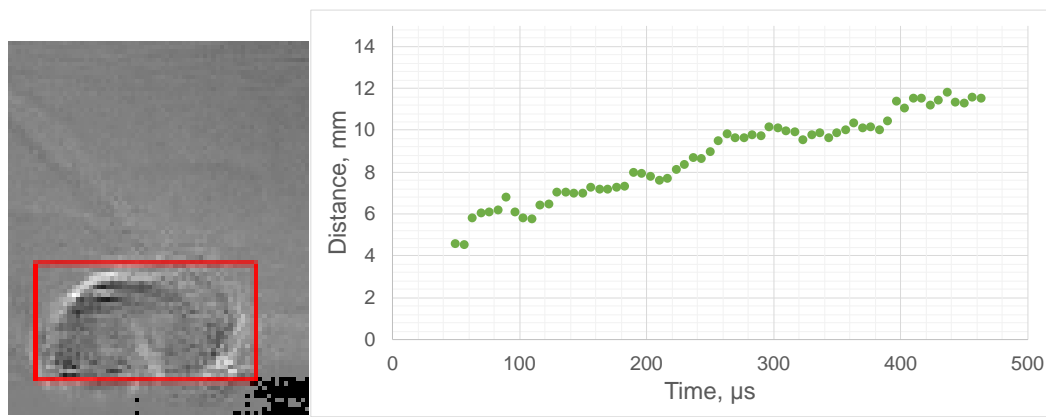

Рис. 11. Распознавание термика от импульсного разряда с помощью CNN и автоматически построенный график его роста.

## 5. ПРИМЕНЕНИЕ ИНФРАКРАСНОЙ ТЕРМОГРАФИИ.

Метод инфракрасной термографии тепловых полей основан на измерении распределения теплового излучения и преобразовании его в карту температуры. Тепловое излучение возникает в твердых телах, жидкостях и газах при температуре выше абсолютного нуля вследствие колебания атомов или вращательно-колебательного движения молекул [63]. Как известно, объекты могут поглощать, отражать или пропускать энергию излучения. Для описания этих физических процессов вводятся соответствующие коэффициенты. ИК излучение лежит в диапазоне спектра электромагнитных волн от 0.75 мкм до 1000 мкм между видимым светом и радиоволнами. В свою очередь ИК спектр принято подразделять на коротковолновую (0.75–1.5 мкм), средневолновую (1.5–20 мкм) и длинноволновую области (20–1000 мкм), хотя в литературе встречаются различные разбиения ИК диапазона в зависимости от дальнейшего применения.

Инфракрасная термография – бесконтактный метод измерения и анализа теплового излучения объектов или потоков. Регистрация теплового излучения дает обширную информацию об энергетическом состоянии объекта исследования, что применяется в теплофизике, медицине, геологии, биологии, энергосбережении, и др.  Использование данного метода встречается в таких инженерных приложениях, как регуляция теплоизоляции, дефектоскопия неразрушающий контроль и др. На рис. 12 приведены примеры термограмм: утечки тепла через щели оконной рамы и изображение "энергосберегающей" лампы; тепловизор с диапазоном регистрации 3.7 – 4.8мкм.

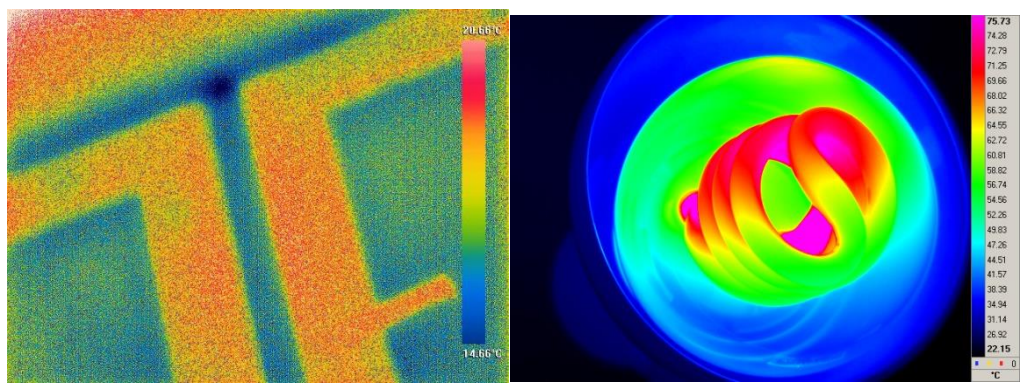

Рис.12. Термограммы теплового поля оконной рамы и светильника с энергосберегающей лампой.

Термография применяется для диагностики психоэмоционального состояния человека в инфракрасном и терагерцовом диапазоне [64 - 66].

Возросший интерес к термографии обусловлен как появлением тепловизоров нового поколения, так и возможностями цифровой обработки, анализа, хранения термографических изображений и фильмов. Полный поток теплового излучения W, регистрируемый тепловизором, равен:

$$W = \tau_a \varepsilon_o W_{АЧТ}(T_o) + \tau_a(1-\varepsilon_o)W_{АЧТ}(T_f) + (1-\tau_a)W_{АЧТ}(T_a).$$

Это общая измерительная формула, используемая в большинстве коммерческих тепловизионных систем. Тепловизор выдает некорректируемое значение температуры объекта, полученное с учетом всех тепловых излучений, принятых детектором.

С помощью термографии может проводиться экспериментальное исследование теплообмена на плоских и рельефных поверхностях с различной геометрией [67, 68].

Несмотря на то, что газовые среды являются прозрачными в ИК области спектра, термография широко применяется для исследования воздушных потоков и их воздействий на поверхность. Ранние известные попытки измерить коэффициенты теплопередачи в высокоскоростном потоке воздуха с помощью термографии были выполнены в гиперзвуковом режиме в аэродинамической трубе [69] В работе [70]. реализована теневая визуализация разреженных потоков в инфракрасном свете (плотность газа в области скачка $\approx 10^{-3}$ кг/м$^3$), что недоступно для визуализации в видимом свете. Так как молекулярный азот не может излучать в инфракрасном диапазоне, то визуализация скачков уплотнения (M=21) связана с преломлением инфракрасного излучения, отраженного от задней стенки рабочей камеры, на скачках уплотнения. Имеет место «прямотеневая» визуализация течения в инфракрасном диапазоне, где коэффициент преломления газов существенно возрастает.

Одно из важных приложений термографии - определение области ламинарно-турбулентного перехода на обтекаемой потоком газа или жидкости поверхности. Проблема контроля ламинарно-турбулентного перехода в газовых средах представляет большой интерес для оптимизации геометрии летательных аппаратов и с точки зрения газодинамики и с точки зрения теплообмена на обтекаемой потоком поверхности. Термография позволили определить зоны ламинарно-турбулентного перехода при обтекании летательных аппаратов потоками воздуха с помощью измерения карт температур на крыльях и лопастях [71-73].

В работах группы из Dryden Flight Research Center описан летный эксперимент по исследованию распределения давления при обтекании плоской пластины в полете на сверхзвуковых скоростях до Maxa 2.0. [74]. Исследован пограничный слой в полете. Цель состояла в том, чтобы определить характеристики перехода пограничного слоя и эффективность использования покрытия поверхности для будущих летных испытаний с использованием ИК термографии. С помощью инфракрасной визуализации были зафиксированы ударная волна, падающая на поверхность в дополнение к определению линии перехода ламинарного пограничного слоя в турбулентный.

Важным направлением термографии является анализ приповерхностных течений жидкости. В литературе преимущественно встречаются работы, посвященные исследованию конфигураций на поверхности раздела газ-жидкость и медленных течений с малыми числами Рейнольдса. ИК излучение поглощается непосредственно на поверхности жидкости. Термография может использоваться для измерений полей температуры жидкости в ламинарных и турбулентных режимах конвективных течений со свободной поверхности жидкости [75]. Получены пространственно-временные характеристики ячеистых и многомасштабных конвективных структур [76].

Исследовались с помощью инфракрасной термографии режимы течения потоков воздуха и жидкости в наклонных трубах на основе тепловых изображений и полей локальных коэффициентов теплоотдачи на нагретой поверхности [77]. В работе [78] ИК термография совместно с методом PIV (цифровое трассирование) используется для анализа структуры свободной жидкой струи, падающей на металлическую пластину в воздухе.

Большое количество работ посвящено применению инфракрасной термографии для исследований теплообмена струйных течений путем регистрации теплового потока от внешней поверхности стенки [79]. Практический интерес представляли задачи контроля и сканирования температурного поля наружных стенок смесителей при течении водного и жидкометаллического теплоносителей [77-81]. Для импактных струй измерения пространственно-временных характеристик турбулентного течения воды обычно проводились через тонкую металлическую стенку (пластину, фольгу) [82-85]. Регистрировались усредненные тепловые поля, полученные за счет теплопередачи исследуемого потока твердой стенке. Основной проблемой, возникающей при измерениях через металлические подложки, является ослабление колебаний температуры испытательной поверхностью. В работе [86] было предложено скомпенсировать ослабление пульсаций температуры с помощью восстановления исходного теплового сигнала с внутренней стороны поверхности путем решения обратного уравнения теплопроводности.

Развитие высокоскоростной термографической техники привело к возможности регистрации достаточно быстропротекающих процессов, в частности - характеристик теплообмена турбулентного течения. В работе [87] ИК съемка используется для визуализации турбулентного течения воды в акриловых круглых трубах с высоким временным разрешением. В работе [88] регистрировались пусковые процессы и динамика незатопленной высокоскоростной струи жидкости на станке гидроабразивной резки. Исследование направлено на получение новых данных о двухфазных потоках в экстремальных условиях, и может быть применено для усовершенствования инженерных гидроструйных конструкций. На рис.13 приведены два термографических изображения развития сверхзвуковой гидроструи при съемке с частотой кадров до 415 Гц, наблюдение ведется с периферийной области струи и ее воздушно-водной оболочки. Скорость истечения струи на оси достигает 270 м/с, (Re ≈ $10^7$).

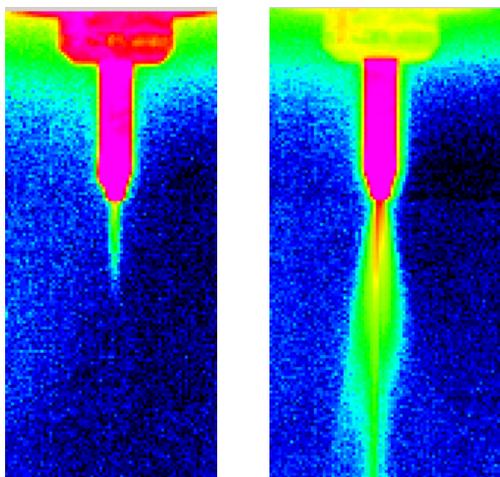

Рис. 13. Термограммы пусковых процессов высокоскоростной гидроструи в период до 0.005 с после запуска.

Свойство воды поглощать инфракрасное (ИК) излучение на субмиллиметровом масштабе позволило предложить метод исследования неизотермических нестационарных турбулентных течений жидкости в пограничном слое [89] на основе ИК термографии (метод ТВПЖ – термография высокоскоростных потоков жидкости). При регистрации через ИК-прозрачное окно метод позволяет визуализировать тепловое излучение из тонкого приповерхностного слоя жидкости. Показано, что через стенку, прозрачную для инфракрасного излучения при тепловизионном исследовании движущейся жидкости могут быть измерены пульсационные, энергетические характеристики неизотермического турбулентного пограничного слоя жидкости с частотным разрешением от 100 Гц. Для модели плоского тройникового соединения выявлено наличие инерционных интервалов энергетических спектров в участках диапазона от 4 до 40Гц. Приложение метода для исследований импактных струй изложено в работах [90-91]

На рис.14 показан пример визуализации переходной области течения импактной струи. Приведены временные развертки пульсации температуры на четырех разных расстояниях (R) от точки торможения.

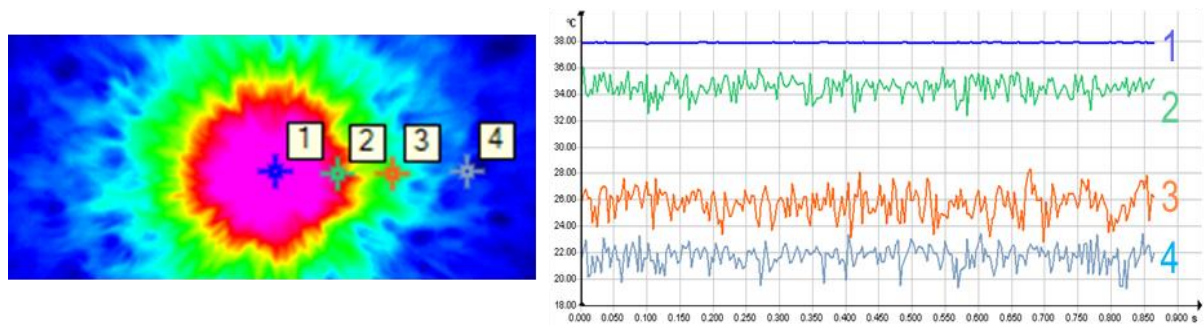

Рис. 14. Пример мгновенной термограммы и временной развертки температуры на четырех разных расстояниях (R) от точки торможения.

## 6. ТЕНЕВОЙ ФОНОВЫЙ МЕТОД

Отдельного рассмотрения заслуживает основанный на физическом явлении рефракции и кросс-корреляционном анализе изображений теневой фоновый метод. При визуализации полей показателя преломления прозрачных неоднородных сред - определяются смещения разделенных в пространстве изображений точек фона, помещенного за исследуемым объектом. Количественные измерения полей плотности возможно в случае двумерного и осесимметричного течения прозрачной среды фотометрическим теневым методом. Величины углов отклонения света связаны с распределением показателя преломления внутри изучаемой неоднородности посредством соотношений Абеля. Например, в работе [92] измерены поля плотности гелия при гиперзвуковом обтекании конуса.

В англоязычной литературе ТФМ известен как Background Oriented Schlieren (BOS). Метод был предложен практически одновременно Майером [93] и Дальцилем [94], тогда же были произведены первые экспериментальные съёмки. Предшественником данного метода можно считать технику спекл-фотографии (СФ), которая развивалась в Институте тепло- и массообмена (ИТМО) им. А. В. Лыкова АН БССР с середины 1980-х годов. Количественная визуализация течений, основанная на спекл-технологиях описана в [95]. Приемы цифровой спекл-фотографии, нашли развитие также в технике PIV, Тальбот-интерферометрии, [96].

Суть теневого фонового метода заключается в сравнении двух изображений одного и того же фона, снятых при отсутствии и при наличии между фотоаппаратом и фоном исследуемого прозрачного объекта с неоднородностями [97; 98]. Фоновый экран должен соответствовать определенным требованиям, для того чтобы при дальнейшей цифровой обработке экспериментальных изображений получить качественные данные с максимальным количеством полезной информации и низким уровнем шума.

Изменение показателя преломления вдоль линии наблюдения в случае съемки фона через объект приводит к несовпадению реперного и «рабочего» изображений (Рис.15). Проанализировав смещение элементов фона на снимках, можно получить количественную информацию об интегральном изменении показателя преломления исследуемой среды вдоль направления наблюдения. Кросс-корреляционные методы цифрового сравнения экспериментальных изображений ранее были существенно разработаны в рамках метода цифровой трассерной анемометрии (PIV).

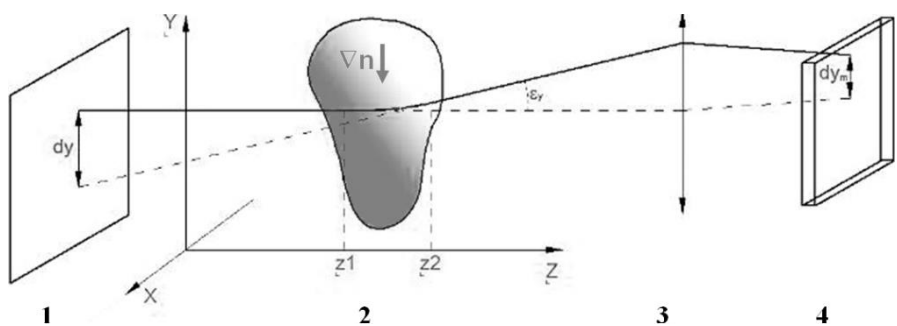

Рис.15. Оптическая схема ТФМ (теневого фонового метода): 1 – фон, 2 – исследуемый объект, 3 – линза/объектив, 4 – цифровой регистратор

Y-компонента отклонения луча, идущего от фона, выражается из закона рефракции следующим образом:

$$\varepsilon_y = \int_{z_1}^{z_2} \frac{1}{n}\frac{\partial n}{\partial y}dz \approx \frac{1}{n_2}\left\langle\frac{\partial n}{\partial y}\right\rangle(z_1 - z_2)$$

Тогда соответствующий элемент фона будет смещён на рабочем изображении относительно реперного на величину:

$$dy = tg\,\varepsilon_y \cdot L_b \approx \varepsilon_y L_b$$

Связь между плотностью однородного газа и ее показателем преломления можно выразить соотношением Гладстона-Дейла:

$$\frac{n-1}{\rho} = G$$

где G – постоянная Гладстона-Дейла. Регистрируемое ТФМ смещение элемента фона на изображении прямо пропорционально градиенту плотности в плоскости, перпендикулярной оптической оси.

$$dy \approx \left\langle \frac{\partial \rho}{\partial y} \right\rangle \frac{G}{\langle n \rangle} l_Z L_b$$

Здесь $L_b$ – расстояние от объекта до фона, $l_z = |z_1 - z_2|$ - толщина исследуемого течения вдоль оптической оси схемы. На смещение изображения влияют все элементы течения вдоль данного оптического луча – фактически, информация об исследуемом потоке усредняется вдоль него. Восстановление непосредственно поля плотности из результатов одноракурсного ТФМ возможно только для течения близкого к двумерному (что редко встречается в эксперименте), либо для осесимметричного течения (на основе использования, обратного преобразования Радона.).

Для получения количественных значений поля плотности в сложных трёхмерных течениях необходима многоракурсная ТФМ-съёмка с последующим восстановлением трёхмерного поля из двумерных изображений с использованием подходов томографической реконструкции. При этом съёмка производится одновременно с нескольких (обычно не менее 5-6) ракурсов. Для стационарных течений съёмка может производиться последовательно с различных углов, одной и той же камерой [99]. Однако даже полученные на основе ТФМ поля смещения (связанные с полями плотности), могут служить полезным источником информации о теплофизических объектах - например, данных о положении и динамике характерных газодинамических структур (разрывов, вихрей и пр.).

В России первые работы с использованием ТФМ были проведены в начале 2000-х в НИУ "МЭИ".

В отличие от теневых методов, ТФМ-визуализация не требует использования оптических элементов, сравнимых по размерам с исследуемым объектом. Это делает его удобным для полевых исследований и экспериментов, и других случаев, когда требуется визуализация течений большого масштаба. Авторы [100] .использовали ТФМ совместно с PIV для комбинированной визуализации течения с числом Маха M=8 в аэродинамической трубе. ТФМ-исследования нестационарных трансзвуковых течений, содержащих ударные волны в канале и на выходе из канала ударной трубы были проведены в МГУ ([101, 102]. На рис.16 приведены 3 ТФМ изображения выхода ударной волны и потока из торца ударной трубы. Поле обзора - до 40 см.

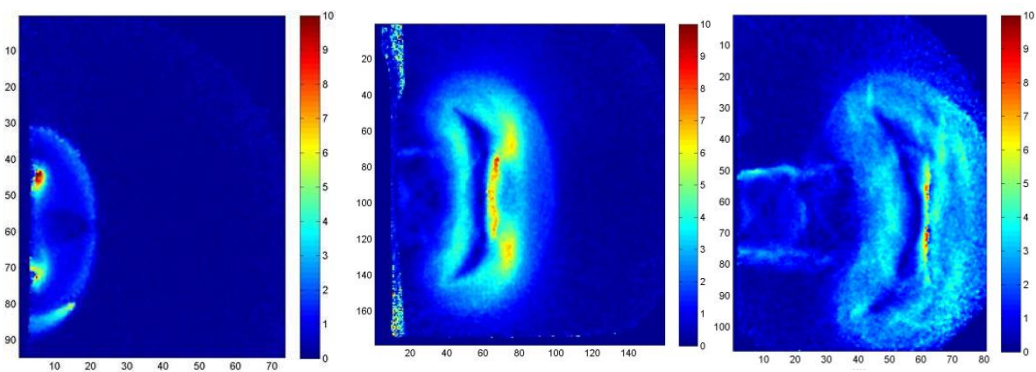

Рис.16. ТФМ изображения сверхзвукового течения при выходе ударной волны в атмосферу.

При исследовании процессов горения ТФМ также может быть широко применён, поскольку метод способен визуализировать не только тепловые потоки и пламена, но также определять концентрации составляющих в различных топливных смесях [103]. ТФМ также применялся для натурных съёмок полномасштабных полевых взрывных испытаний, [104 - 106]. Основным результатом было определение динамики фронта ударной волны, порождённой взрывом. Результаты работ однозначно указывают на то, что ТФМ обладает исключительными возможностями по практическим полевым применениям, хотя при этом не всегда удаётся производить количественные измерения.

Практически с самого начала использования теневого фонового метода были обнаружены значительные погрешности, возникающие при визуализации сильных градиентов плотности (в частности, ударных волн) данным методом. Результаты многочисленных работ показывают, что количественное определение скачка плотности на фронте ударной волны посредством классической схемы ТФМ представляет сложную проблему. Эффект, регистрируемый на фронтах ударных волн, зачастую не коррелирует с физическими параметрами потока. В работах [107], [101] было показано, что данная проблема фактически обусловлена выходом детектируемой величины рефракции за рамки чувствительности метода. Из-за сильного преломления света на фронте ударной волны отклонённый луч может выйти за пределы оптической схемы и не быть зарегистрированным.

На сегодняшний день большинство работ, связанных с ТФМ, рассматривает течения и процессы, происходящие в газе. Однако по своим принципам ТФМ также применим к исследованию процессов в прозрачных жидких и аморфных средах. Так, метод использовался для регистрации внутренних волн в объёме воды [108] ) и тепловых процессов в воде и плексигласе ( МЭИ 2008). На рис. 17 приведены изображения полей смещения при движении конвективных термиков в плоском сосуде.

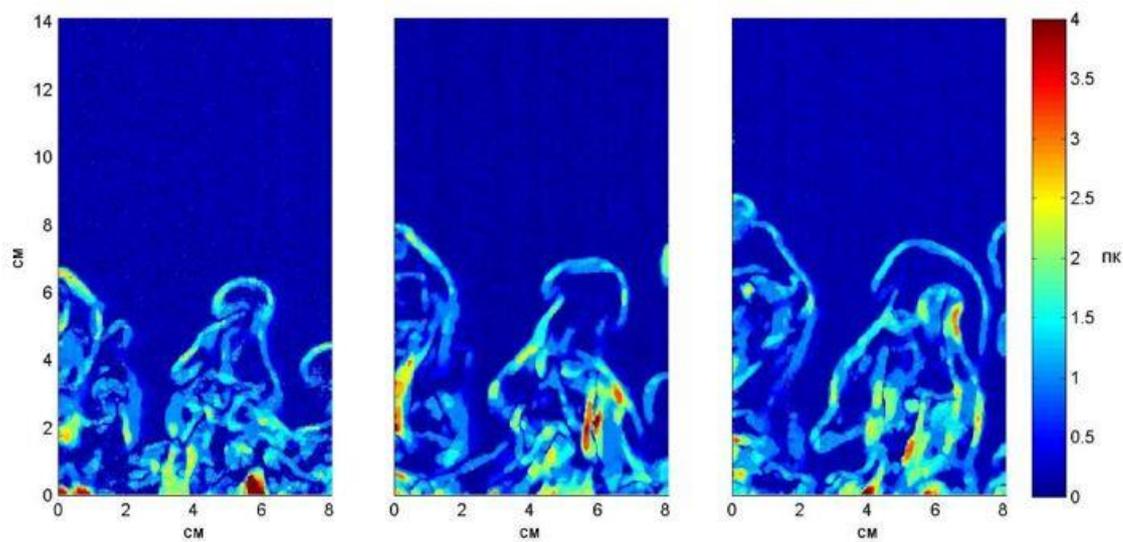

Рис.17. ТФМ изображения конвективных потоков в воде.

ТФМ занимает нишу классических качественных теневых методов, выигрывая у них за счёт большей простоты применения [109] там, где не требуется высокое пространственное разрешение. Большая простота аппаратной части ТФМ по сравнению с другими методами делает более доступной многоракурсную ТФМ-визуализацию.

При всех преимуществах ТФМ очевиден ряд недостатков [110, 111]. Алгоритмы кросс-корреляции изображений усредняют по размеру окна опроса, ТМФ всегда обеспечивает меньшее разрешение изображения, чем сопоставимые традиционные теневые схемы. Повышение чувствительности требует большего окна опроса и, таким образом, понижается разрешение. Потеря разрешения наиболее заметна при таких объектах, как ударные волны и границы раздела сред. В то время, как традиционные теневые методы дают готовое изображение в реальном времени (без обработки) ТФМ требует обработки. Существенны проблемы с фокусировкой одновременно на объекте и фоне. Метод также чувствителен к вибрациям. Одним из недостатков, усложняющим количественные измерения с помощью теневого фонового метода, является непараллельность световых лучей, зондирующих исследуемый объём. Это приводит к неравномерности пространственного масштаба ТФМ-полей в зависимости от положения фиксируемых неоднородностей вдоль оптического луча. Попытки скомпенсировать данный недостаток путём введения в оптическую систему большой собирающей линзы, согласованной с объективом регистрирующей камеры, отменяют одно из существенных достоинств метода, т.к. ограничивает размеры исследуемой неоднородности размерами главного оптического элемента.

## 7. ЦИФРОВОЕ ТРАССИРОВАНИЕ

Движение элементов газа и жидкости можно визуализировать, вводя в поток меченые, окрашенные частицы, струйки дыма, нити, шелковинки, и. т. д. Такой метод визуализации - метод трассирования - является одним из старейших способов прямой визуализации течения. При интегральной регистрации следа частицы в газе метод трассирования позволяет наблюдать траектории частиц, линии тока. Регистрируя с экспозицией δt движение трассирующих частиц в потоке, можно получить изображения отрезков пути δs, пройденные частицами за δt. Величина v=δs/δt представляет собой среднюю скорость частицы на этом отрезке. Цифровая обработка результатов трассирования развилась в специальное направление обеспечивающее реконструкцию динамики трехмерных полей скорости - метод PIV – (particle image velocimetry).

Метод цифрового трассирования (анемометрии по изображениям частиц, международное название метода Particle Image Velocimetry) основан на статистическом анализе смещений изображений частиц, движущихся вместе с исследуемым течением, визуализирующих поток за короткий интервал времени при регистрации этих частиц в выделенной плоскости с помощью оптического ножа. Во времени оптический нож (лазерное излучение) сформируется в виде двух коротких импульсов некоторым интервалом времени между импульсами излучения. Итогом измерения PIV метода являются мгновенные поля скорости потока. Множество обзоров и монографий посвящено описанию модификаций и применений метода [112, 114]

В России метод активно развивался с 90-х годов в институте Теплофизики им. С.С. Кутателадзе . СО РАН. В обзоре [15] дан анализ истории и современных тенденций развития метода анемометрии по изображениям частиц в приложениях к аэродинамическим установкам. Авторами рассматриваются основы метода анемометрии, варианты его реализации, история и современное состояние аппаратных средств.

Появление термина PIV относят к 80-м годам, когда он был выделен как частный случай метода лазерной спеклометрии LSV (Laser Speckle Velocimetry) [115-117]. базирующегося на оптическом преобразовании Фурье яркостных картин.

К планарным модификациям метода относятся Particle Image Velocimetry (PIV), Micro Particle Image Velocimetry (Micro PIV), Particle Tracking Velocimetry (PTV) и др.

К объемным методам исследования кинематической структуры потока относятся Stereo Particle Image Velocimetry (Stereo PIV), Tomographic Particle Image Velocimetry (Tomographic PIV) и др.

Метод PTV практически идентичен с методом PIV. Итогом измерения PTV метода также являются мгновенные двухкомпонентные поля скорости. Но в отличие от PIV метода вектор скорости измеряется по перемещениям отдельных трассеров в потоке, а не группы. Метод PTV применяется в случае, если плотность образов частиц очень мала. Обработка изображений также производится с помощью корреляционных алгоритмов.

Одним из важнейших преимуществ метода является отсутствие значительного возмущающего влияния на поток. К ограничениям PIV можно отнести конечные размеры трассирующих частиц, вследствие чего трассеры не всегда точно следуют за потоком. Особенно это касается областей сильных градиентов и разрывов. Размер используемых частиц ограничивает размер элементарной области, при этом использование более мелких частиц приводит к влиянию броуновского движения на их положение. Трассеры для газовых сред могут быть жидкими каплями размером 1-100 мкм; как правило, используются различные натуральные и синтетические масла. Для высокоскоростных потоков используются твёрдые частицы, чаще всего из оксида титана и алюминия. Они могут иметь меньшие размеры (порядка сотен нм), и за счёт этого лучше следовать потоку.

Область применения PIV-метода в частности включает в себя фундаментальные научные исследования, направленные на изучение динамики и масштабов вихревых структур в потоках жидкости и газа, [118-120]. В работах [121-122] продемонстрирована визуализация методом PIV профилей скорости за взрывными (ударными) волнами, образующимися при подрыве взрыве проволочки. С помощью 8-импульсной лазерной системы получены последовательные изображения полей скорости за фронтом сферической взрывной волны. Отмечается, что на фронте волны зафиксировано значительное уширение профиля скорости, по-видимому, обусловленное преимущественно инерционным запаздыванием жидких трассирующих частиц. Этот эффект, отмечаемый и в других работах, ограничивает применимость метода к исследованиям полей течений при взрывах.

Визуализация сверхзвуковых потоков с помощью PIV остаётся сложной проблемой в связи с неравномерной плотностью частиц на изображениях. В связи с этим визуализация сверхзвуковых течений требует особенно тщательного подбора трассирующих частиц, механизмов засева потока, алгоритмов и параметров обработки изображений PIV.

Другой вопрос, тесно связанный с предыдущим – корректировка данных PIV в потоках с большими градиентами скорости. Из-за запаздывания частиц, данные в таких течениях могут существенно отличаться от истинных. Однако в некоторых случаях удаётся учесть эти погрешности и восстановить истинное поле скорости газа [123]. В работе [124] методом PIV измерены нестационарные поля скоростей, возникающие при развитии потоков за ударными (взрывными) волнами, инициируемыми импульсным поверхностным скользящим разрядом в воздухе. Плазменные листы (наносекундные разряды, скользящие по поверхности диэлектрика) были инициированы на стенах прямоугольной камеры. Распределения скоростей потока за этими волнами показал, что вложение импульсной энергии является однородным вдоль разрядных каналов плазменного листа, в то время как интегральная видимая интенсивность свечения плазмы уменьшается в направлении распространения канала. На рис.18 приведено поле скоростей за плоской ударной волной (слева) и за цилиндрической ударной волной, инициированной наносекундным плазменным каналом.

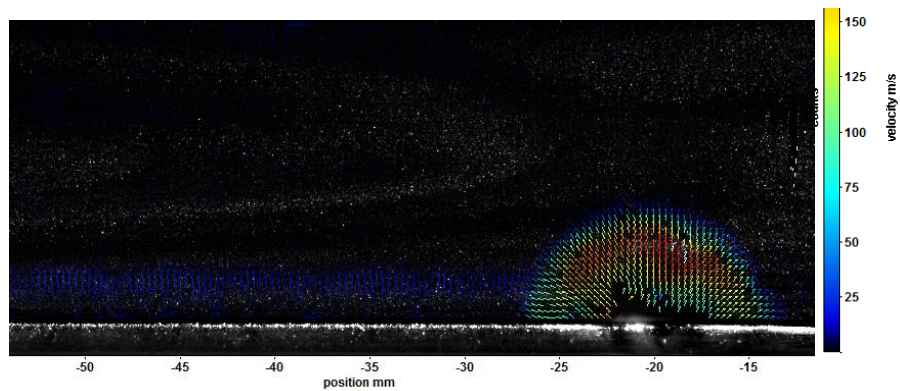

Рис.18. Поле скоростей за ударной волной, инициированной поверхностным скользящим разрядом.

Увеличивается в последние годы число приложений PIV для задач исследования микропотоков [125-126]. В последней работе представлены результаты применения метода micro-PIV для визуализации структуры потока в капле воды, расположенной на стеклянной подложке (рис. 19). В различные моменты испарения капли получены двухкомпонентные поля скорости в разных сечениях капли по высоте.

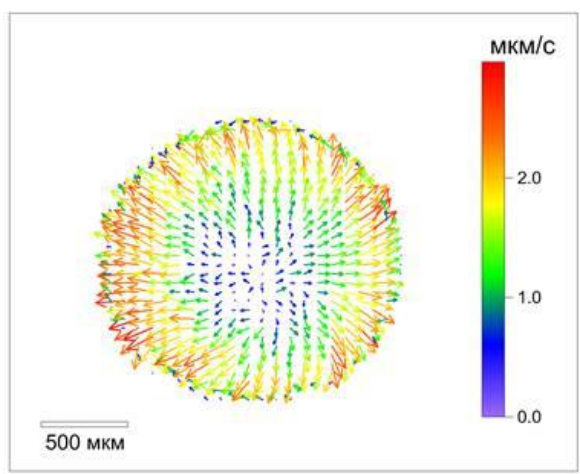

Рис. 19. Мгновенное поле скорости в испаряющейся капле. Сечение – 45 мкм от подложки.

В работе [127] проведена PIV визуализация и исследование течения, развивающегося при выходе ударной волны из канала ударной трубы в атмосферу (рис.20).

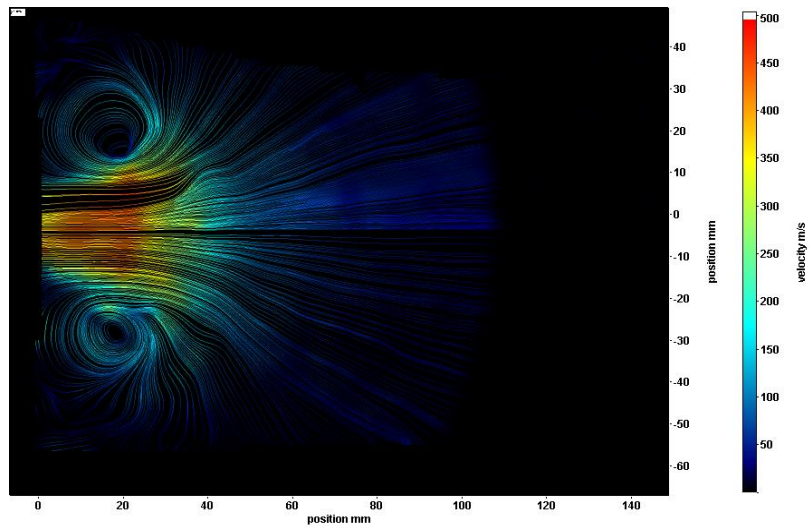

Рис. 20. Линии тока течения за ударной волной, выходящей из канала
(цвет кодирует скорость потока).

В последние годы благодаря распространению кросс-корреляционных алгоритмов обработки изображений появилось значительное количество работ по использованию беззасевного трассирования – слежения за структурными элементами, маркерами, присутствующими в самом потоке. В этой методике "шлирен PIV" используются естественные неоднородности и преломляющие турбулентные вихри в потоке в качестве виртуальных «затравочных частиц», на которых проводится измерение скорости [128-131].
 Пример визуализации неоднородностей пограничного слоя при высокоскоростной теневой съемке приведен на рис. 21. Результат кросс-корреляционной обработки подобных изображений может быть использован как нижняя оценка скорости газа в пограничном слое.

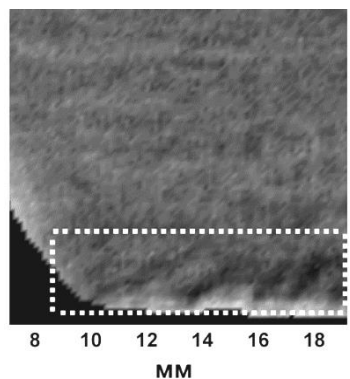

Рис.21. Кадр высокоскоростной теневой съёмки течения в рабочей секции ударной трубы

Предложенный в "Казанском научном центре РАН" [132] метод дымовой визуализации основан на цифровой обработке видеозаписей визуализации дыма, записанных на световом листе. Потоки газа засеваются генераторами аэрозоля, которые следуют за потоком газа так же т, как и в PIV, но из-за более высокой концентрации они выглядят на изображении как не освещенные отдельные частицы, а дым с непрерывной интенсивностью. Метод позволяет использовать относительно примитивное оборудование для измерений динамики двухкомпонентных векторных полей скорости с частотой около 10 кГц.

Одним из беззасевных методов трассерной визуализации является новый метод трассирования тепловыми точками (ТТТ), основанный на термографической визуализации пограничного слоя жидкости [133]. Программой кросс-корреляции измеряется смещение точек равной температуры (в градациях серого) на двух соседних термограммах. На рис.22 приведено усредненное по 100 кадрам поле скоростей потока в пограничном слое воды на основе метода ТТТ. Использовалась высокоскоростная съемка тепловизором импактной затопленной неизотермической струи через окно, прозрачное для инфракрасного излучения. В центре – область ламинарного течения.

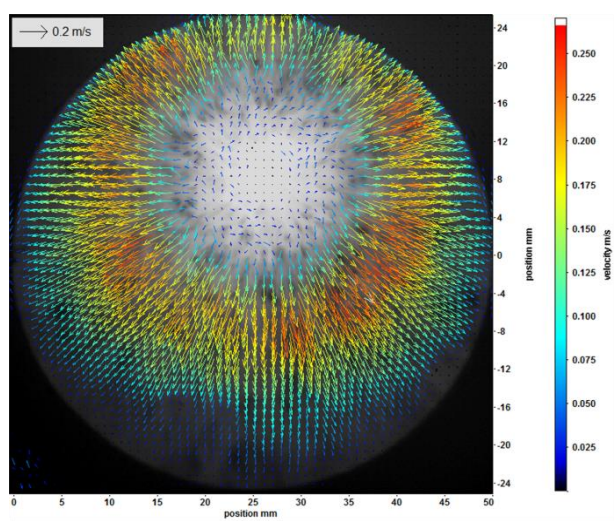

Рис.22. Метод беззасевного трассирования тепловыми точками. Поле скоростей в пограничном слое импактной затопленной струи

# 8.МЕТОДЫ ВИЗУАЛИЗАЦИИ ПОВЕРХНОСТНОГО ТЕЧЕНИЯ

Гидродинамическое течение на обтекаемой твердой поверхности, может быть визуализировано с использованием специально разработанных методов. Они позволяют детектировать области перехода ламинарного пограничного слоя к турбулентному, визуализировать зоны отрыва потока, области пересечения скачков уплотнения с поверхностью, распределение теплофизических параметров по поверхности и др. При этом исследуются процессы в переходной области потока – в пограничном слое на поверхности. В переходной области течения (динамический пограничный слой) происходит изменение скорости от нуля на стенке до некоторого конечного значения во внешнем потоке. Можно также выделить тепловой (температурный) пограничный слой, образующийся в случае несовпадения температуры поверхности и температуры газа. При протекании на стенке химической реакции или вдуве образуется концентрационный (диффузионный) пограничный слой. Тепловой поток от газа к модели можно вычислить, если измерить скорость нагрева поверхности модели с известными теплофизическими свойствами.

Оптические панорамные методы различаются способом преобразования температуры поверхности в оптический сигнал: люминесцентные преобразователи температуры, жидкие кристаллы, тепловидение. Использование специальных методов визуализации поверхностного течения позволяет получить картины, созданные распределением скорости, температуры, давления, пристеночными касательными напряжениями в пограничном слое. Визуализация поверхностного течения осуществляется, как правило, следующим образом: на обтекаемую поверхность модели наносится специальная краска, жидкая пленка или другое покрытие, реагирующее на локальные параметры течения. Затем регистрируется картина, созданная распределением давления, температуры, скорости, пристеночными касательными напряжениями на этой поверхности. В качестве покрытий, используемых для визуализации поверхностного течения, используются: жидкие кристаллы, покрытия для инфракрасной термографии, покрытия, чувствительные к давлению (баротропные покрытия), и к температуре (термоиндикаторные краски), саже-масляные покрытия, вязкое масло, некоторые специальные покрытия, используемые в особых случаях. Методы дают как количественные, так и качественные изображения полей параметров на поверхности модели без внесения возмущений в поток (бесконтактные методы диагностики). Полученное изображение (фильм) течения на поверхности регистрируется на цифровой носитель во время эксперимента (или после него). Существенна проблема восстановления визуализирующих свойств покрытия во время эксперимента; для ее решения применяются специальные методы.

Возможность измерения интенсивности и частоты света открыло дорогу созданию методов, в которых интенсивность света связана с измеряемым параметром (давлением, температурой, напряжением трения и т.д.). В первую очередь речь идет о методах люминесцентных преобразователей давления и температуры.

В конце 70-х годов внимание специалистов ЦАГИ (Центральный АэроГидродинамический Институт, г. Жуковский) привлекли работы доцента Ленинградского технологического института И. Захарова, который исследовал тушение люминесценции органических красителей (люминофоров) молекулами кислорода. Они использовали это явление для измерения распределения давления воздуха (в состав которого входит кислород) на поверхности модели в аэродинамических трубах. Саму идею ученые запатентовали в 1980 г., а реализующее ее первое чувствительное к давлению покрытие (слой силикагеля с адсорбированными в нем молекулами органического люминофора) - в 1981 г. Это покрытие - авторы назвали его бароиндикатором - получилось несовершенным, но с его помощью уже можно было увидеть поле давления. К концу 1989 г. было создано несколько различных типов покрытий, чувствительных к давлению (их назвали люминесцентными преобразователями давления ЛПД).

Использование явления люминесценции в оптических методах обладает тем преимуществом по сравнению с использованием рассеянного и отраженного света, что благодаря спектральному сдвигу и временной задержке позволяет отстроиться от падающего (возбуждающего) света и, тем самым, повысить точность измерения интенсивности света, несущего полезную информацию.

Сегодня определение распределения давления на обтекаемой поверхности с использованием покрытий, чувствительных к локальному давлению (PSP - Pressure Sensitive Paint) - один из основных современных панорамных бесконтактных методов диагностики поверхностных течений в теплофизике.

Для визуализации распределения давления поверхность модели покрывается несколькими специальными слоями краски; один из слоев содержит флуоресцирующее вещество, с оптической активностью, зависящей от парциального давления кислорода. Физическое явление, лежащее в основе визуализации полей давления - тушение люминесценции органических красителей кислородом воздуха. Возбужденный светом соответствующей длины волны, люминофор может излучать свет (люминесценция) или терять энергию, передав ее молекулам кислорода (тушение люминесценции). Доля теряемой энергии пропорциональна концентрации кислорода в полимере и его подвижности. Согласно закону Генри, концентрация кислорода в полимерном слое прямо пропорциональна его парциальному давлению над поверхностью полимера. Таким образом, квантовый выход люминесценции обратно пропорционален парциальному давлению кислорода. Если концентрация кислорода в воздухе постоянна, то эффект тушения люминесценции может быть использован для измерения давления воздуха. Люминесцентный преобразователь температуры (ЛПТ) – это покрытие, которое под действием возбуждающего излучения люминесцирует с интенсивностью, зависящей от его температуры.

Первые публикации об использовании метода были опубликованы сотрудниками ЦАГИ [134]. Аналогичные зарубежные работы появились только через 4–5 лет. В работе [135] полимерный быстрый ЛПД исследовался в лабораторной ударной трубе, а в работе [136] приведены результаты измерений давления и температуры в гиперзвуковой ударной трубе университета Колспан. Наиболее полные обзоры по теме представлены в монографии [137] и в книге [138]. В этом новом издании описываются краски, чувствительные к давлению и температуре (PSP и TSP), для панорамных измерений поверхностного давления и температуры в аэродинамике и механике жидкости. Книга включает в себя последние достижения в области составов красок, приведены результаты стационарных и нестационарных измерений в различных аэродинамических установках, включая сверхзвуковые и гиперзвуковые аэродинамические трубы. Описываются технические аспекты, включая калибровку, освещение, обработку изображений, - Обсуждаются неопределенность измерений PSP и TSP (Temperature Sensitive Paints- покрытия чувствительные к температуре).

В работах [139 -141] описаны методы и результаты измерения полей давления и температур и визуализации линий поверхностного течения и напряжений трения сдвига. На рисунке 23 представлено поле давления на поверхности модели, полученное с помощью люминесцентных преобразователей давления.

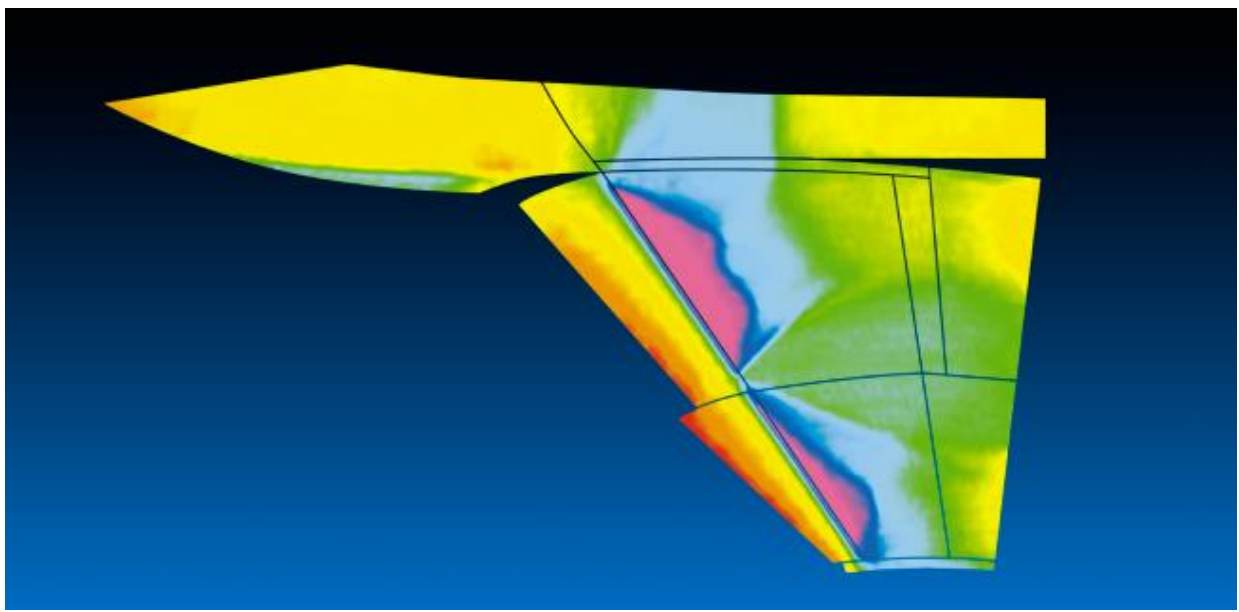

Рис.23. Распределение давления на пластине (http://www.tsagi.ru/research/measurements/lpd2.jpg)

Поверхностные линии тока или направление потока у поверхности модели можно визуализировать, если нанести на поверхность масляную пленку. Оптически контрастные (люминесцентные) твёрдые частицы добавляются в плёнку вязкого масла, и по меньшей мере два распределения этих частиц регистрируются на поверхности модели в потоке через заданный интервал времени с помощью цифровой камеры. Обработка этих изображений с помощью корреляционного анализа (аналогично тому, как это делается в методе PIV) даёт вектора смещения частиц. Для расчета поля напряжений сдвига необходимо также знать толщину масляной пленки и динамическую вязкость масла в каждой точке поверхности модели [142, 143]. Метод получил широкое распространение при изучении отрывных течений. На рисунке 24 представлена полученная методом масляных покрытий картина визуализации линий тока и напряжения трения.

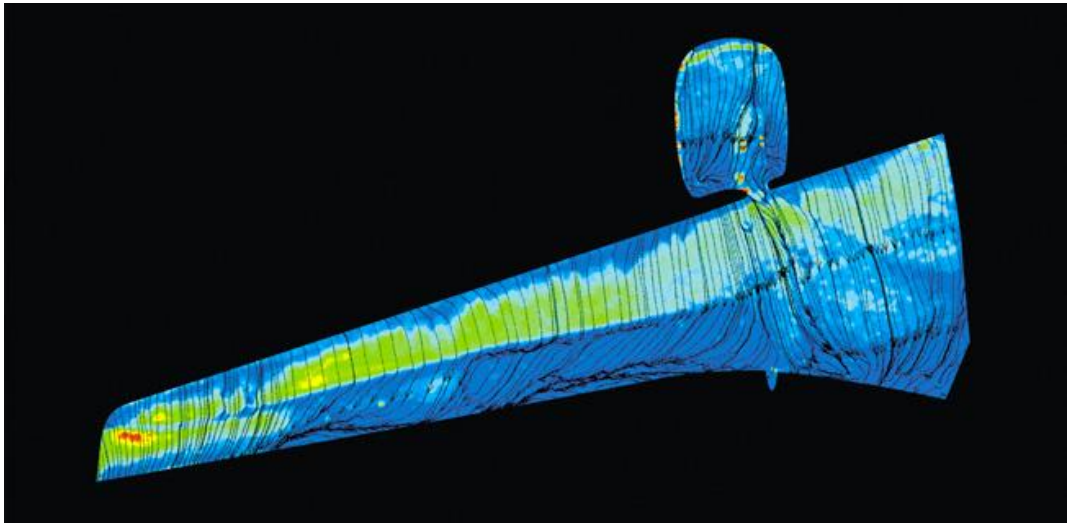

Рис.24 Метод «густого» масла. Обтекание крыла пассажирского самолета
http://www.tsagi.ru/research/measurements/ofi/measurements_maslo.jpg

Поверхностная интерферометрия в масляных пленках - метод измерения сопротивления обшивки летательного аппарата при обтекании потоком. Слой прозрачного масла, нанесенного на поверхность, обтекаемую воздушным потоком, становится тоньше в области возникновения напряжения сдвига. Локальное изменение толщины слоя может быть измерено с использованием явления интерференции в тонких пленках. Интерференция возникает при освещении поверхности, покрытой масляной пленкой переменной толщины излучением лазера. Интерференция в тонких пленках происходит при взаимодействии света, отраженного от воздушно-масляной поверхности раздела и поверхности раздела масло – твердая поверхность. Для получения количественной информации о сопротивлении обшивки измеряется величина смещения полос.

Простой, но весьма эффективный метод визуализации структуры фронта детонационной волны используется в газодинамике горения для анализа структуры фронта детонации [143]. Существование пересжатых и недосжатых участков на плоском ударном фронте волны детонации ведет к возникновению неоднородностей. Периодические неоднородности во фронте детонации в виде линий двойного и тройного пересечения участков волн оставляют следы на закопченной стеклянной пластинке помещенной в торце трубы. Это явление и сегодня используется для исследования структуры фронта детонации, масштаба неоднородности фронта.

Жидкокристаллические покрытия используются как в воздушных, так и в водяных потоках. Высокая пространственная разрешающая способность, малые времена отклика, многоцветность, обратимость, сравнительно низкая стоимость позволяют использовать жидкие кристаллы для визуализации распределения температуры и напряжения сдвига на поверхности модели, линий тока в пространственных течениях в широком диапазоне режимов обтекания [144]. Термочувствительные жидкие кристаллы меняют цвет в зависимости от температуры. За пределами диапазона температур они становятся прозрачными на черной поверхности. При повышении температуры цвет меняется, и граница области цвета указывает положение изотермы на поверхности. Диапазон температур, измеряемых с помощью жидкокристаллических покрытий – примерно от - 40 до 280° С. Однако в одном опыте визуализируется диапазон в несколько градусов, что означает фактически визуализацию одной изотермы при большом разбросе температур на поверхности.

# 9.ВИЗУАЛИЗАЦИЯ ДАННЫХ ЧИСЛЕННОГО МОДЕЛИРОВАНИЯ: ИМИТАЦИЯ РЕЗУЛЬТАТОВ ПАНОРАМНОГО ЭКСПЕРИМЕНТА.

Характерная особенность современного этапа визуализации теплофизических потоков – стремительное сближение численных и экспериментальных панорамных изображений теплофизических полей. Такое сближение стало возможным в связи с внедрением цифровых технологий в методы регистрации и анализа потоков с одной стороны, и расширением возможностей численного моделирования, с другой стороны. На основе сравнения с данными экспериментальной визуализации потоков проводится верификация моделей и алгоритмов численных расчетов. С другой стороны, результаты эксперимента расшифровываются и уточняются на основе данных численного эксперимента.

На рис.25 приведено полученное методом визуализации импульсным объемным разрядом мгновенное изображение стадии нестационарного процесса дифракции плоской ударной волны в прямоугольном канале 48х24 мм на прямоугольном препятствии на стенке канала. На экспериментальное изображение нанесен газодинамический расчет двумерного течения (изолинии плотности, уравнения Навье- Стокса). Расчет помогает расшифровать детали экспериментального изображения.

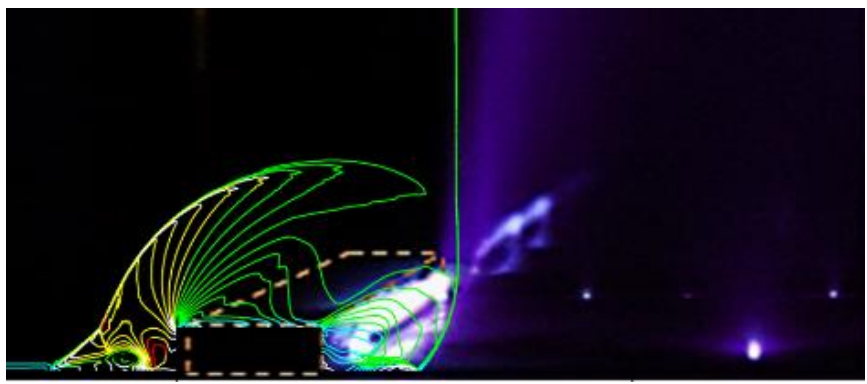

Рис.25. Свечение разряда и расчетное поле плотности в канале с уступом при движении ударной волны.

Данные численного моделирования визуализируются при возможности в виде имитации результатов эксперимента (численные интерферограммы, теневые изображения, люминесцентные покрытия, PIV, ТФМ и др.).

Сегодня бОльшая часть результатов расчетных исследований сверхзвуковых газодинамических процессов представляется в виде численных теневых картин. Наиболее эффективно это происходит в структурированных течениях – с разрывами, выраженными неоднородностями, вихревыми структурами.

На рис.26 приведены расчетное и экспериментальное изображения полей плотности сложного газодинамического квазидвумерного течения в ударной трубе после взаимодействия ударной волны с импульсным объемным разрядом с предионизацией от плазменных электродов. Сравнение экспериментальных теневых изображений с результатами расчетов на основе уравнений Эйлера и Навье–Стокса – позволило путем решения обратной задачи рассчитать значение энерговклада в поток [145-147].

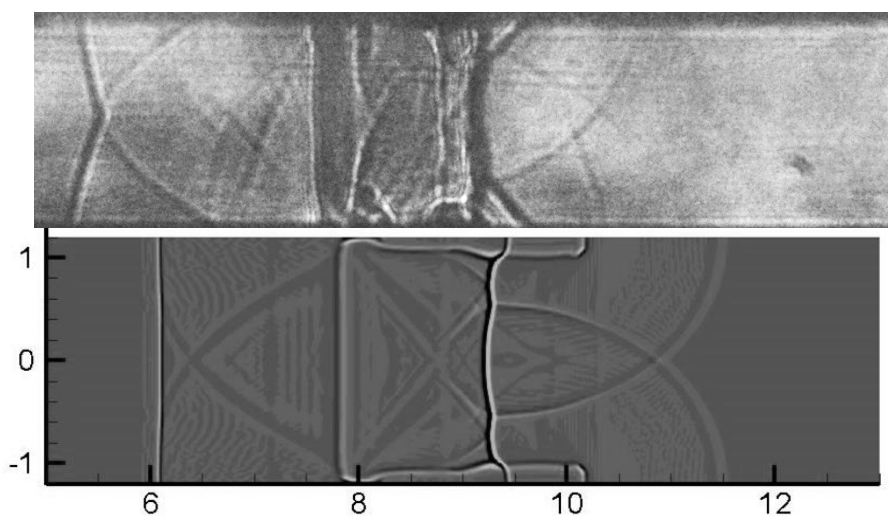

Рис.26. Сравнение расчетных и экспериментальными теневых изображений.

В 90-х гг. были опубликованы первые компьютерные интерферограммы двумерных течений [148, 149].

В книге [20] рассматриваются подходы и приводятся многочисленные примеров визуального представления решения ряда задач газовой динамики, связанных с расчетами течений невязкого и вязкого сжимаемого газа, содержащих слабые и сильные газодинамические разрывы. Приведены данные визуализации сверхзвуковых струйных и отрывных течений, полученные при экспериментальных и численных исследованиях.

На рис.27 приведены изображения сверхзвукового течения при выходе ударной волны из прямоугольного канала, полученного теневым фоновым методом (слева) и численным моделированием соответствующего потока при тех же параметрах.

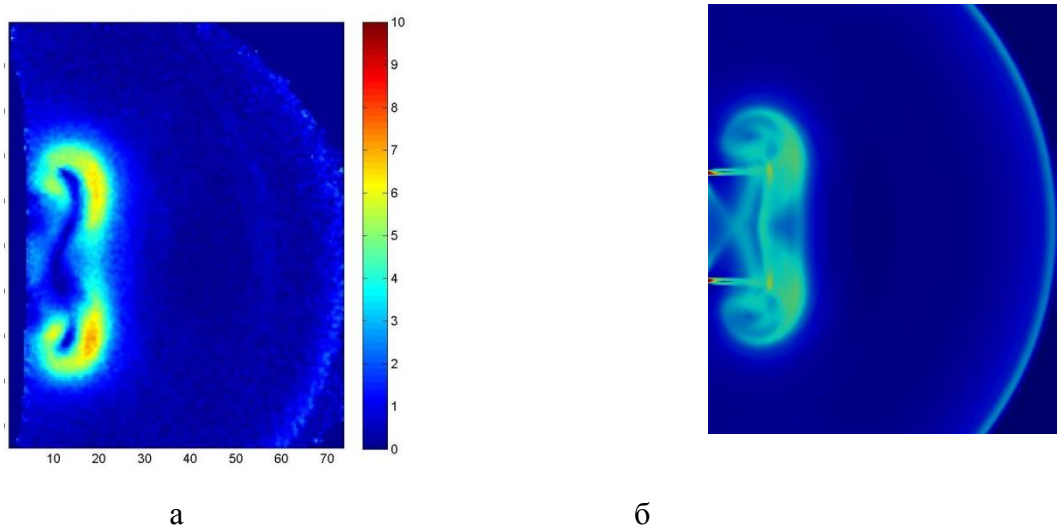

а б

Рис.27. Сверхзвуковое двумерное течение с ударной волной и спутным потоком. Теневой фоновый метод – поле смещений (а) и численное моделирование (б).

На рис.28 приведен результат визуализации численного моделирования того же течения – представление поля скоростей – имитация метода цифрового трассирования (PIV).

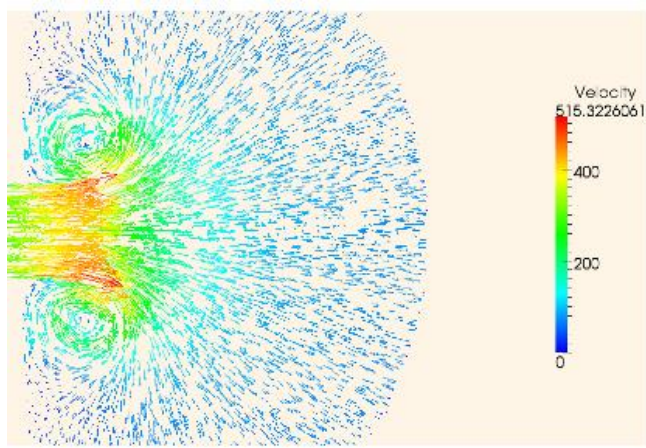

Рис.28. Численное моделирование – поле скоростей сверхзвукового течения с ударной волной.

**Благодарности.**

# Список литературы

1. Mach, E., Salcher, P. "Photographische Fixirung der durch Projectile in der Luft eingeleiteten Vorgänge". Sitzungsber. Kaiserl. Akad. Wiss., Wien, Math.-Naturwiss. Cl. (in German). 95 (Abt. II): 764–780. 1887. Doi 10.1002/andp.18872681008.
2. Etienne-Jules Marey, ''Des mouvements de l'air lorsqu'il rencontre des surfaces de differentes formes,'' Comptes Rendus des Seances de l'Acad´emie des Sciences 131 (July 16, 1900): 160–63.
3. Ван Дайк. Альбом течений жидкости и газа // М.: Мир, 1986.
4. Yang W.J. (ed.) Handbook of Flow Visualization // NY.: Hemisphere Publishing Corporation. 1989.
5. Merzkich W. Flow Visualization // 2nd edition. NY.:Academic Press. 1987.
6. Мишин Г.И. Оптические методы исследований в баллистическом эксперименте. // Л.: Наука. 1979.
7. Климкин В.Ф., Папырин А.Н., Солоухин Р.Н. Оптические методы регистрации быстропротекающих процессов // Н.: Наука. 1980.
8. Глотов Г.Ф., Майкапар. Аэротермодинамика летательных аппаратов в фотографиях. 2003, ЦАГИ, Жуковский.
9. Panigrahi P. K., Muralidhar K., Schlieren and Shadowgraph Methods in Heat and Mass Transfer Springer, 2012.
10. Settles G.S. Schlieren and Shadowgraph Techniques. Visualizing Phenomena in Transparent Media Springer. 2001.
11. Ronald J. Adrian, Jerry Westerweel, Particle Image Velocimetry (Cambridge Aerospace ), 2010.
12. Дубнищев Ю. Н., Арбузов В. А., Белоусов П. П., Белоусов П. Я. Оптические методы исследования потоков. Новосибирск: Сибирское университетское издательство, 2003, 418 С.
13. Белозёров А.Ф. Оптические методы визуализации потоков. Казань: издательство КГТУ,2007,747 С.
14. Базылев Н. Б., Фомин Н. А. Количественная визуализация течений, основанная на спекл-технологиях Минск : Беларуская навука, 2016. – 392 С.
15. Вавилов В.П. Инфракрасная термография и тепловой контроль. 2-е издание, доп. М. Издательский дом Спектр. 2009. С. 544.
16. Современные оптические методы исследования потоков / под ред. Б. С. Рикевичюса. – М.: Оверлей, 2011.
17. Бильский А.В., Гобызов О.А., Маркович Д.М. История и тенденции развития метода анемометрии по изображениям частиц для аэродинамического эксперимента (обзор) // Теплофизика и аэромеханика. 2020. Т. 27. № 1. С. 1-24.
18. Пимштейн, В. Г. Аэроакустические взаимодействия в турбулентных струях / Moscow, 2010. – 88 с. – ISBN 9785922112321.
19. Большухин М. А. и др. Актуальные задачи развития экспериментальной базы для верификации CFD кодов при использовании в атомной энергетике. Труды НГТУ им. Р.Е. Алексеева, 2013, Т 2(99) стр 117-125.
20. Волков К.Н., "Визуализация данных физического и математического моделирования в газовой динамике" Москва : Физматлит, 2018. - 356 с..
21. А. Е. Бондарев, В. А. Галактионов, В. М. Чечёткин, "Анализ развития концепций и методов визуального представления данных в задачах вычислительной физики", Ж. вычисл. матем. и матем. физ., 51:4 (2011), 669–683.
22. Krehl P., Engemann S. August Toepler // Shock Waves. 1995. V. 5. Is. 1–2. P. 1–18.
23. Литвиненко Ю.А., Козлов В.В., Грек Г.Р., Литвиненко М.В., Шмаков А.Г. Диффузионное горение круглой микроструи водорода при до- и сверхзвуковой скорости истечения. Доклады Российской академии наук. Физика, технические науки. 2020. Т. 494. № 1. С. 25-30.
24. Зудов В.Н., Третьяков П.К., Тупикин А.В. Воспламенение и стабилизация оптическим разрядом гомогенного горения в высокоскоростной струе, 2016, том 8, номер: 2, 24 – 36 С.
25. Васильев Л.А. Теневые методы // Наука. 1968.
26. Kleine H., Hiraki K., Maruyama H. et al. High-speed time-resolved color schlieren visualization of shock wave phenomena // Shock Waves. 2005. V. 14. Is. 5-6. P. 333–341.
27. Иншаков С.И., Родионов А.Ю., Ширин А.С., Шехтман В.Н. Одиннадцатая Международная научно-техническая конференция «Оптические методы исследования потоков», Москва, 27 — 30 июня 2011 г. интерферометр для одновременной регистрации двух интерферограмм с ортогональным направлением сдвига.
28. Букин В.В., Гарнов С.В., Малютин А.А., Стрелков В.В. труды института общей физики им. а.м. прохорова интерферометрическая диагностика фемтосекундной лазерной микроплазмы в газах, 2011, Том 67, С. 3-3.
29. Дубнищев Ю.Н. и др. Оптическая диагностика конвективных структур, индуцированных нестационарными граничными условиями в вертикальном слое воды. // Научная визуализация, 2018, том 10, номер 4, 134 – 144 С.
30. Wei W., Lia X., Wu J., Yang Z., Jia S., Qiu A. Interferometric and schlieren characterization of the plasmas and shock wave dynamics during laser-triggered discharge in atmospheric air // Physics of Plasmas, Vol. 21, No. 8, 2014.
31. Hargather J., Settles S. A review of recent developments in schlieren and shadowgraph techniques // Meas. Sci. Technol 2017. V. 28. N. 4.
32. Кандауров А.А., Сергеев Д.А., Ермакова О.С., Троицкая Ю.И. Исследование теневыми методами механизмов генерации брызг при ветроволновом взаимодействии // Научная визуализация Т.9 N3 103 – 107
33. Алферов В.И., Бушмин А.С. Электрический разряд в сверхзвуковом потоке воздуха // ЖЭТФ, 1963, Т. 44, В. 6, стр. 1775-1779.
34. Алферов В.И., Окерблом Т.И., Саранцев А.И. Экспериментальное исследование вихревого течения около крыльев малого удлинения и круглых конусов при числе Маха, равном двум // Изв. АН СССР МЖГ, 1967, № 5, С. 113-121.
35. Алферов В.И. Исследование структуры электрического разряда большой мощности в высокоскоростном потоке воздуха // Изв. АН СССР, МЖГ, 2004, № 6, стр. 163-175.
36. Алферов В.И. К вопросу определения плотности потока при визуализации вихревых жгутов методом высоковольтного разряда // Тр. ЦАГИ. 1972. Вып. 1421. С. 13-2.
37. Алферов В.И., Дмитриев Л.М. Электрический разряд в потоке газа при наличии градиентов плотности // ТВТ, 1985, Т. 23, № 4, стр. 677-682.
38. Nishio M., Sezaki S., Nakamura H. Visualization of Flow Structure Around a Hypersonic Re-entry Capsule Using the Electrical Discharge Method // J. of Visualization, Vol.7, No.2, 2004.
39. Nishio M., Nakamura H., Sezaki Sh., Manabe K. Flowfield Around Space Plane Traveling at Mach 10 (Comparison of Visualization and Calculation) // The 10th International Symposium on Flow Visualization, Aug.26-29, 2002, Kyoto, Japan.
40. Jagadeesh G., Srinivasa Rao B.R., Nagashetty K., Reddy N.M., Reddy K.P.J. Electrical discharge technique for three-dimensional flow field visualisation in hypersonic shock tunnel // J. Flow Visualisation and Image Processing, Vol. 4, N0. 1, pp. 51-57, 1997.
41. Jagadeesh G., Srinivasa Rao B.R., Nagashetty K., Reddy K.P.J., Viren M. Visualization studies around spiked blunt cones using electrical discharge technique at Mach 5.75 // The 10th International Symposium on Flow Visualization, Aug. 26-29, 2002, Kyoto, Japan.
42. Chen X., Sha X., Wen S., Lu H.B., Ji F. Visualization of three dimension shock wave in hypersonic gun tunnel using electric discharge. Proceedings 18th International Symposium on Flow Visualization. Zurich, Switzerland, 2018.
43. Ieshkin A.E, Danilov A.V, Chernysh V.S, Ivanov I.E, Znamenskaya I.A. Visualization of supersonic flows with bow shock using transversal discharges. Journal of Visualization, Vol. 22, pp. 741–750, 2019.
44. Znamenskaya I. A., Kuli-zade T. A. , Kulikov V. N., Perminov S. P. Transonic 3d non-stationary flow visualization using pulse transversal discharge // Journal of Flow Visualization and Image Processing. — 2011. — Vol. 18, no. 3. — P. 214-224.
45. Знаменская И. А., Кули-Заде Т. А. Визуализация неустойчивости тороидального вихря импульсным объемным разрядом // Доклады Академии наук. — 1996. — Т. 348, № 5. — С. 617–619.
46. Знаменская И. А., Иванов И. Э., Крюков И. А., Кули-Заде Т. А. Импульсный объемный разряд с предионизацией в двумерном газодинамическом потоке // Журнал экспериментальной и теоретической физики. — 2002. — Т. 122, № 6. — С. 1198–2006.

47. Знаменская И. А., Татаренкова Д. И., Кули-заде Т. А. Наносекундная ионизация области обтекания прямоугольного уступа высокоскоростным потоком // Письма в Журнал технической физики. — 2020. — Т. 46, № 1. — С. 5–7.
48. Li G. et al. Image Processing Techniques for Shock Wave Detection and Tracking in High Speed Schlieren and Shadowgraph Systems. // Journal of Physics: Conference Series, vol. 1215, 2019.
49. Edla D. et al. Advances in Machine Learning and Data Science: Recent Achievements and Research Directives. Springer, 2018.
50. Ye S. et al. A flow feature detection method for modeling pressure distribution around a cylinder in non-uniform flows by using a convolutional neural network. Scientific Reports, vol. 10, 2020.
51. Canny J. A. Computational Approach to Edge Detection. IEEE Transactions on Pattern Analysis and Machine Intelligence, vol. PAMI-8(6), p. 679-698, 1986.
52. Srisha Rao M., Jagadeesh G. Visualization and Image Processing of Compressible Flow in a Supersonic Gaseous Ejector. // Journal of the Indian Institute of Science, vol. 93, no. 1, 2013.
53. Fujimoto T.R., Kawasaki T., Kitamura K. Canny-Edge-Detection/Rankine–Hugoniot-Conditions Unified Shock Sensor for Inviscid and Viscous Flows // J. Comput. Phys., vol. 396, pp. 264–279, 2019.
54. Brunton S. L. et al. Machine Learning for Fluid Mechanics. Annual Review of Fluid Mechanics, vol. 52, pp. 477-508, 2020.
55. Dehghan Manshadi M. et al. Speed Detection in Wind-tunnels by Processing Schlieren Images. IJE TRANSACTIONS A: Basics, vol. 29, no. 7, pp. 962-967, 2016.
56. Colvert B. et al. Classifying vortex wakes using neural networks. Bioinspiration & Biomimetics, vol 13, no. 2, 2018.
57. Harel R., Rusanovsky M., Fridman Y., Shimony A., Oren G. Complete Deep Computer-Vision Methodology for Investigating Hydrodynamic Instabilities. High Performance Computing. ISC High Performance 2020. Lecture Notes in Computer Science, Vol 12321, p. 61-80, 2020.
58. Ott C., Pivot C., Dubois P., Gallas Q., Delva J., Lippert M., Keirsbulck L. Pulsed jet phase-averaged flow field estimation based on neural network approach. Experiments in Fluids, Vol. 62, No. 79, 2021.
59. Kutz J. Deep learning in fluid dynamics. // Journal of Fluid Mechanics, Vol. 814, pp 1-4, 2017.
60. Znamenskaya I. A., Doroshchenko I.A. Edge detection and machine learning for automatic flow structures detection and tracking on schlieren and shadowgraph images // Journal of Flow Visualization and Image Processing. — 2021. — Vol. 28, no. 4. — P. 1–26.
61. Знаменская И. А., Дорощенко И. А., Сысоев Н. Н., Татаренкова Д. И. Результаты количественного анализа высокоскоростной теневой съемки течений в ударной трубе при помощи машинного зрения и обучения // Доклады Академии наук. — 2021. — Т. 497, № 1. — С. 16–20.
62. Znamenskaya, I.A. et al. Edge detection and machine learning approach to identify flow structures on schlieren and shadowgraph images. Proceedings of the 30th International Conference on Computer Graphics and Machine Vision. CEUR Workshop Proceedings, vol. 2744, pp. 1–14, 2020.
63. Вавилов В.П. Инфракрасная термография и тепловой контроль. 2-е издание, доп. М. Издательский дом Спектр. 2013. – P. 544.
64. Chernorizov A., Isaychev S., Znamenskaya I., Koroteyeva E., Khakhalin A., Shishakov V. Remote Detection of Human Emotional States by Facial Areas // International Journal of Psychophysiology. – 2018. – 131. – S. 85.
65. Berlovskaya E.E., Isaychev S.A., Chernorizov A. M., Ozheredov I.A., Adamovich T.V., Isaychev E.S., Cherkasova O.P., Makurenkov A.M., Shkurinov A.P., Varaksin A.N., Gatilov S.B., Kurenkov N.I., Manaenkov A.E. Diagnosing Human Psychoemotional States by Combining Psychological and Psychophysiological Methods with Measurements of Infrared and THz Radiation from Face Areas // Psychology in Russia: State of the Art. – 2020. – 13. – №. 2. – P. 64-83.
66. Carlomagno G.M., Cardone G. Infrared thermography for convective heat transfer measurements // Experiments in fluids. – 2010. – 49. – №. 6. – P. 1187-1218.
67. Leontiev A.I., Kiselev N.A., Burtsev S.A., Strongin M M., Vinogradov Y.A. Experimental investigation of heat transfer and drag on surfaces with spherical dimples // Experimental Thermal and Fluid Science. – 2016. – 79. – P. 74-84.
68. Kiselev N.A., Leontiev A.I., Vinogradov Yu.A. et al. Effect of large-scale vortex induced by a cylinder on the drag and heat transfer coefficients of smooth and dimpled surfaces // International Journal of Thermal Sciences. – 2019. – 136. – P. 396-409.
69. Thomann H., Frisk B. Measurement of heat transfer with an infrared camera //International Journal of Heat and Mass Transfer. – 1968. – 11. – №. 5. – P. 819-826.
70. Базовкин В. М., Ковчавцев А. П., Курышев Г. Л., Маслов А. А., Миронов С. Г., Хотяновский Д. В., Царенко А. В., Цырюльников И. С. Численное и экспериментальное исследование обтекания двумерного угла сжатия гиперзвуковым потоком. // Журнал Вестник НГУ. Серия Физика. 2007. Том 2, № 1. С. 3–9.
71. Simon B., Filius A., Tropea C., Grundmann S. // Experiments in Fluids. – 2016. – 57. – №. 5. – P. 93.
72. Raffel M., Merz C.B. Differential Infrared Thermography for Unsteady Boundary-Layer Transition Measurements // AIAA journal. – 2014. – 52. – №. 9. – P. 2090-2093.
73. Richter K., Schulein E. Boundary-layer transition measurements on hovering helicopter rotors by infrared thermography // Experiments in fluids. – 2014. – 55. – №. 7. – P. 1755.
74. Banks D.W., Frederick M.A., Tracy R.R., Matisheck J.R., Vanecek N.D. In-flight boundary-layer transition on a large flat plate at supersonic speeds. 15th International Symposium on Flow Visualization ISFV15 – Minsk / Belarus – 2012 0-62.
75. Бердников В.С., Гришков В.А., Ковалевский К.Ю., Марков В.А. Тепловизионные исследования ламинарно-турбулентного перехода в Рэлей-Бенаровской конвекции //Автометрия. – 2012. – Т. 48. – №. 3. – С. 111-120.
76. Иваницкий Г. Р., Деев А. А., Хижняк Е. П. Структуры на поверхности воды, наблюдаемые с помощью инфракрасной техники // Успехи физических наук. Т. 175, N 11. 2005 С. 1207-1216.
77. Hetsroni G., Mewes D., Enke C., Gurevich M. et al. Heat transfer to two-phase flow in inclined tubes // International Journal of Multiphase Flow. – 2003. – 29. – №. 2. – P. 173-194.
78. **Ошибка! Источник ссылки не найден.**
79. Carlomagno G.M., Ianiro A. Thermo-fluid-dynamics of submerged jets impinging at short nozzle-to-plate distance: A review // Experimental thermal and fluid science. – 2014. – 58. – P. 15-35.
80. Judd K.P., Smith G.B., Handler R.A., Sisodia A. The thermal signature of a low Reynolds number submerged turbulent jet impacting a free surface //Physics of Fluids. – 2008. – 20. – №. 11. – 115102.
81. Кашинский О.Н., Лобанов П.Д., Курдюмов А.С., Прибатурин Н.А. Экспериментальное моделирование течения жидкометаллического теплоносителя в Т-образном смесителе // Журнал технической физики. – 2016. – Т. 86. – № 5. – С. 145-147.
82. Зайцев Д.К., Смирнов Е.М., Колесник Е.В., Большухин М.А., Будников А.В., Свешников Д.Н. Расчетно-экспериментальное исследование температурных пульсаций в тройниковом соединении с косым впрыском // Сборник докладов «Проблемы применения и верификации CFD кодов в атомной энергетике». – 2018. – С. 104-105.
83. Bol'shov L., Pribaturin N., Kashinsky O., Lobanov P., Kurdyumov A. Experimental Study of Mixing Fluid Flows with Different Temperatures in a T-Junction // Journal of Applied Mechanics and Technical Physics. – 2020. – 61. – №. 3. – P. 368-376.
84. Nakamura H., Shiibara N., Yamada S. Quantitative measurement of spatio-temporal heat transfer to a turbulent water pipe flow // International Journal of Heat and Fluid Flow. – 2017. – 63. – P. 46-55.
85. Roux S., Fenot M., Lalizel G., Brizzi L.-E., Dorignac E. Evidence of flow vortex signatures on wall fluctuating temperature using unsteady infrared thermography for an acoustically forced impinging jet // International journal of heat and fluid flow. – 2014. – 50. – P. 38-50.
86. Nakamura H. Measurements of time-space distribution of convective heat transfer to air using a thin conductive film // Fifth International Symposium on Turbulence and Shear Flow Phenomena. – Begel House Inc., 2007. – P. 1906–1914.
87. Shiibara N., Nakamura H., Yamada S. Visualization of turbulent heat transfer to a water flow in a circular pipe using high-speed infrared thermography //Journal of Flow Visualization and Image Processing. – 2013. – 20. – №. 1-2.

88. Znamenskaya I.A., Koroteeva E.Yu., Shirshov Ya.N., Novinskaya A.M., Sysoev N.N. High speed imaging of a supersonic waterjet flow //Quantitative InfraRed Thermography Journal. – 2017. – Т. 14. – №. 2. – С. 185-192.
89. Большухин М.А., Знаменская И.А., Фомичев В.И. Метод количественного анализа быстропротекающих тепловых процессов через стенки сосудов при неизотермическом течении жидкости // Доклады Академии наук. издательство Наука М. 2015. – 465. – № 1.– С. 38– 42.
90. Koroteeva E., Shagiyanova A., Znamenskaya I., Sysoev N. Time-resolved thermographic analysis of the near-wall flow of a submerged impinging water jet // Experimental Thermal and Fluid Science. — 2021. — P. 110 – 264.
91. Znamenskaya I., Koroteeva E., Shagiyanova A. Thermographic analysis of turbulent non-isothermal water boundary layer // Journal of Flow Visualization and Image Processing. — 2019. — Vol. 26, no. 1. — P. 49–56.
92. Храмцов П.П. и др. Диагностика полей плотности фотометрическим теневым методом при гиперзвуковом обтекании конуса в легкогазовой баллистической установке // Журнал технической физики, 2019, том 89, вып. 10, 1506-1514.
93. Meier G., (2002). Computerized background-oriented schlieren. Exp. Fluids 33 (1), pp.181–187.
94. Dalziel S.B., Hughes G.O., Sutherland B.R., (2000). Whole-field density measurements by "synthetic schlieren." Exp. Fluids 28 (4), pp.322–335.
95. Базылев Н. Б., Фомин Н. А. Количественная визуализация течений, основанная на спекл-технологиях – Минск : Беларуская навука, 2016. – 392 с.
96. Khramtsov P.P., Penyazkov O.G., Shatan I.N. Temperature measurements in an axisymmetric methane–air flame using Talbot images. Exp Fluids 56, 31 (2015).
97. Raffel M. Background-oriented schlieren (BOS) techniques. Exp Fluids 56, 60 (2015).
98. Скорнякова Н.М. Теневой фоновый метод и его применения. В кн. Современные оптические методы исследования потоков / под ред. Б. С. Рикевичюса. – М.: Оверлей, 2011. С 93-107.
99. Hazewinkel J., Maas L. R. M., Dalziel S. B. Tomographic reconstruction of internal wave patterns in a paraboloid. Experiments in Fluids, vol. 50, no. 2, pp. 247–258, Jul. 2010.
100. Kirmse T., Agocs J., Schröder A., Martinez Schramm J., Karl S., Hannemann, K., (2011). Application of particle image velocimetry and the background-oriented schlieren technique in the high-enthalpy shock tunnel Göttingen. Shock Waves 21 (3), pp.233–241.
101. Glazyrin F.N., Znamenskaya I.A., Mursenkova I.V., Sysoev N.N., Jin J., (2012). Study of shock-wave flows in the channel by schlieren and background oriented schlieren methods. Optoelectron. Instrum. Data Process. 48 (3), pp.303–310.
102. Глазырин Ф., Знаменская И., Коротеева Е., Мурсенкова И., Сысоев Н. Использование теневого фонового метода для исследования нестационарного потока с ударной волной // Научная Визуализация – 2013. – Т. 5 – № 3 – 65–74с.
103. Tillmann W., Abdulgader M., Rademacher H., Anjami N., Hagen L., (2014). Adapting of the Background-Oriented Schlieren (BOS) Technique in the Characterization of the Flow Regimes in Thermal Spraying Processes. J. Therm. Spray Technol. 23 (1-2), pp.21–30.
104. Mizukaki T., Wakabayashi K., Matsumura T., and Nakayama K., (2014). Background-oriented schlieren with natural background for quantitative visualization of open-air explosions. Shock Waves 24 (1), pp.69–78, 2014.
105. Герасимов С.И., Трепалов Н.А. Регистрация воздушных ударных волн с помощью теневого фонового метода. Научная визуализация 2017, 4,9 с.1-12.
*106.* Сысоев Н. Н., Знаменская И.А. Новые возможности цифровых технологий анализа изображений при испытаниях на полигонах. Известия Российской академии ракетных и артиллерийских наук. — 2020. — Т. 112, № 2. — С. 114.
107. Znamenskaya I.A., Vinnichenko N.A., Glazyrin F.N., (2012). Quantitative measurements of the density gradients on the flat shock wave by means of background oriented schlieren. (ISFV, Minsk, Belarus), p. 060.
108. Hazewinkel J., Maas L.R.M., Dalziel S.B., (2011). Tomographic reconstruction of internal wave patterns in a paraboloid. Exp. Fluids 50 (2), pp.247–258.
109. Hayasaka K., Tagawa Y. Mobile visualization of density fields using smartphone background-oriented schlieren. Exp Fluids 60, 171 (2019).
110. Hargather M.J., Settles G.S. (2012) A comparison of three quantitative schlieren techniques. Opt Lasers Eng 50(1):8–17.
111. Fisher T.B., Quinn M.K., Smith K.L. An experimental sensitivity comparison of the schlieren and background-oriented schlieren techniques applied to hypersonic flow // Measurement Science and Technology, 2019, V. 30, N. 6.
112. Adrian R. J. Particle Image Velocometry / Adrian R. J., Westerweel J. Particle Image Velocometry. Evolution and recent trends of particle image velocimetry for an aerodynamic experiment (review) // Cambridge University Press, 2010. – 586 p.
113. Raffel M., Willert C.E., Scarano F., Kähler C., Wereley S.T., Kompenhans J. 2018. Particle image velocimetry, a practical guide 3rd ed. Springer Int. Publishing. 669 p.
114. Scarano F. 2012. Tomographic PIV: principles and practice // Meas. Sci. Technol. Vol. 24, No. 1. P. 012001-1- 012001-28.
115. Grant, I. Particle image velocimetry: A review. Arch. Proc. Inst. Mech. Eng. C J. Mech. Eng. Sci. 1997, 211, 55–76.
116. Н. Б. Базылев, Н. А. Фомин. Количественная визуализация течений, основанная на спекл-технологиях Минск : Беларуская навука, 2016. 392с.
117. Adrian R. J. Twenty years of particle image velocimetry // Exp. in Fluid s. – 2005. – Vol . 39. – P. 159–169.
118. Mariani R., Kontis K., (2010). Experimental studies on coaxial vortex loops. pp.126102. Physics of Fluids 22(12).
119. Скорнякова Н.М., Сычев Д.Г., Вараксин А.Ю., Ромаш М.Э. (2015). Визуализация вихревых структур методом анемометрии по изображениям частиц. Научная Визуализация 7 (3), pp.15–24.
120. Koroteeva E.Y., Znamenskaya I.A., Glazyrin F.N., Sysoev N.N. Numerical and experimental study of shock waves emanating from an open-ended rectangular tube // Shock Waves. — 2016. — Vol. 26, no. 3. — P. 269–277.
121. Murphy M.J., Adrian R.J., (2010). PIV space-time resolution of flow behind blast waves. Exp. Fluids 49 (1), pp.193–202.
122. Murphy M.J., Adrian, R.J., (2011). PIV through moving shocks with refracting curvature. Exp. Fluids 50 (4), pp.847–862.
123. Koroteeva E., Mursenkova I., Liao Y., Znamenskaya I. Simulating particle inertia for velocimetry measurements of a flow behind an expanding shock wave // Physics of Fluids. — 2018. — Vol. 30, no. 1. — P. 011702.
124. Глазырин Ф. Н., Знаменская И. А., Мурсенкова И. В. и др. Анализ однородности энерговклада при развитии каналов плазменного актуатора на основе цифрового трассирования // Письма в Журнал технической физики. — 2016. — Т. 42, № 2. — С. 16–22.
125. Bin Yang, Yuan Wang, Wen Bo He "Application of Micro-PIV on the Microscale Flow and a Modified System Based on Ordinary 2-D PIV." Advanced Materials Research 346 (2011): 657–63.
126. Ягодницына А.А., Бильский А.В., Кабов О.А. Визуализация течения в испаряющейся капле на подложке с помощью метода MICRO-PIV // Научная визуализация, 2016, т.8, N 2, c. 53-58.
127. Koroteeva E. Y., Znamenskaya I. A., Glazyrin F. N., Sysoev N. N. Numerical and experimental study of shock waves emanating from an open-ended rectangular tube // Shock Waves. — 2016. — Vol. 26, no. 3. — P. 269–277.
128. Dennis R. Jonassen, Gary S. Settles, Michael D. Tronosky, Schlieren "PIV" for turbulent flows, Optics and Lasers in Engineering, Volume 44, Issues 3–4, 2006, Pages 190-207.
129. Michael John Hargather Michael James Lawson Gary S. Settles Leonard M. Weinstein. Seedless Velocimetry Measurements by Schlieren Image Velocimetry 2011 AIAA Journal 49(3):611-620.
130. Nematollahi O., Samsam-Khayani H., Kim K.C., Nili-Ahmadabadi M., Yoon S.Y A novel self‐seeding method for particle image velocimetry measurements of subsonic and supersonic flows Scientific Reports. 2020. Т. 10. № 1. С. 10834.
131. Bryan E. Schmidt , Wayne E. Page, Jeffrey A. Sutton Seedless Velocimetry in a Turbulent Jet using Schlieren Imaging and a Wavelet-based Optical Flow Method AIAA 2020-2207 Session: Advancements in Planar, Volumetric, and High-Speed Imaging Techniques.
132. Михеев Н.И., Душин Н.С. 2016. метод измерения динамики векторных полей скорости турбулентного потока по видеосъемке дымовой визуализации // Приборы и техника эксперимента. № 6. С. 114-122.

133. Koroteeva E. Y., Znamenskaya I. A., Ryazanov P. A. Velocity-field measurements in a fluid boundary layer based on high-speed thermography // Doklady Physics. — 2020. — Vol. 65, no. 3. — P. 100–102.
134. Borovoy V., Bykov A., Mosharov V., Orlov A., Radchenko V., Fonov S. Pressure Sensitive Paint Application in Shock Wind Tunnel. // 16th ICIASF Congress. - Dayton, Ohio, July 1995, ICIASF 95 record, 1995. P. 34.1–34.4.
135. Hubner J.P., Carroll B.F., Schanze K.S., Ji H.F. Pressure sensitive Paint Measurements in a Shock Tube. // Experiments in Fluids, 2000. - V. 28. - № 1. – P. 21–28.
136. Hubner J.P., Carroll B.F., Schanze K.S., Ji H.F., Holden M.S. Temperature- and Pressure-Sensitive Paint Measurements in Short-Duration Hypersonic Flow. // AIAA Journal, 2001. - V. 39. - № 4. – P. 654-659.
137. Liu T., Sullivan J.P. Pressure and Temperature Sensitive Paints // Springer 2005, 328c.
138. Liu T., Sullivan J.P., Asai K., Klein C., Egami Y. 2nd ed. 2021. Springer.
139. Мошаров В.Е. Люминесцентные методы исследования течений газа на поверхности.// ПТЭ. №1. 2009. С.5-18.
140. Mosharov V., Radchenko V. PSP/TSP activity in TsAGI // 15th International Symposium on Flow Visualization (ISFV15) proceedingss, Minsk, Belarus, June 25-28, 2012, CD, ISFV15-009, pp.1-4.
141. Mosharov V., Radchenko V., Tsipilev N. Particle image surface flow visualization and skin-friction measurement // 31st Congress of the International Council of the Aeronautical Sciences, ICAS 2018 : 31, Belo Horizonte, 2018.
142. Мошаров В. Е., Радченко В. Н., Ципилев Н. С. Визуализация течения на поверхности по изображениям частиц: шаг к измерению поверхностного трения // Модели и методы аэродинамики : Шестнадцатая Международная школа-семинар, Евпатория, 2016– С. 118-119 ЦАГИ.
143. Щелкин К.И., Трошин Я.К. Газодинамика горения. М. 1963.
144. Zharkova G.M., Kovrizhina V.N., Petrov A.P., Shapoval, E.S., Mosharov V.E. and Radchenko V.N. Visualization of boundary layer transition by shear sensitive liquid crystals.// Proceedings PSFVIP-8: 2011 Moscow, Russia. No. 113. - P. 1-5.
145. Znamenskaya I. A., Koroteev D. A., Lutsky A. E. Discontinuity breakdown on shock wave interaction with nanosecond discharge // Physics of Fluids. — 2008. — Vol. 20. — P. 056101–1–056101–6.
146. Четверушкин Б. Н., Знаменская И. А., Луцкий А. Е., Ханкасаева Я. В. Численное моделирование взаимодействия и эволюции разрывов в канале на основе компактной формы квазигазодинамических уравнений // Математическое моделирование. — 2020. — Т. 32, № 5. — С. 44–58.
147. Koroteeva E., Znamenskaya I., Orlov D , Sysoev N.Shock wave interaction with a thermal layer produced by a plasma sheet actuator // Journal of Physics D - Applied Physics. — 2017. — Vol. 50, no. 8. — P. 085204.
148. Tamura Y., Fujii K. Visualization for computational fluid dynamics and the comparison with experiments, Paper AIAA-90-3031 (1990).
149. Fursenko A. A., Sharov D. M., Timofeev E. V., Voinovich P. A. Numerical Simulation of Shock Wave Interactions with Channel Bends and Gas Nonunoformities. // Computers Fluids Vol. 21, No. 3, pp. 377-396, 1992.